\documentclass[a4paper,fleqn,usenatbib,useAMS]{mn2e}

\usepackage{graphicx}	
\usepackage{amsmath}	
\usepackage{amssymb}	
\usepackage{multicol}        
\usepackage{bm}		
\usepackage{pdflscape}	
\usepackage{natbib}
\usepackage{times}
\usepackage{calc}
\usepackage{rotating}
\usepackage{lscape}
\bibpunct{(}{)}{;}{a}{}{,} 
\usepackage{longtable}
\usepackage{enumitem} 
\usepackage{hyperref}
\usepackage{pifont}
\usepackage{tikz}

\def\oversim#1#2{\lower0.5pt\vbox{\baselineskip0pt \lineskip-0.5pt
     \ialign{$\mathsurround0pt #1\hfil##\hfil$\crcr#2\crcr\sim\crcr}}}

\usepackage[T1]{fontenc}
\usepackage{ae,aecompl}

\usepackage{mathptmx}

\def\apj {{ApJ}}

\def\aap {{A\&A}}
\def\mnras {{MNRAS}}

\def\pasp  {{PASP}}

\def\aj {{ApJ}}

\def\apjs {{ApJs}}

\title[Outflows in Planetary Nebulae]{
  Kinematical Investigation of Possible Fast Collimated Outflows in
  Twelve Planetary Nebulae}

\author[Rechy-Garc\'\i a et al.]{J.S.\ Rechy-Garc\'\i a$^1$, M.A.\ Guerrero$^1$, S.\ Duarte Puertas$^1$, Y.-H.\ Chu$^2$, 
\newauthor{J.A.\ Toal\'a$^3$, \& L.F.\ Miranda$^1$} \\
\\
$^1$ Instituto de Astrof\'isica de Andaluc\'ia, IAA-CSIC,
Glorieta de la Astronom\'\i a s/n, 18008 Granada, Spain. \\
$^2$Institute of Astronomy and Astrophysics, Academia Sinica 
(ASIAA), Taipei 10617, Taiwan, Republic of China \\
$^3$ Instituto de Radioastronom\'\i ıa y Astrof\'\i sica (IRyA), 
UNAM Campus Morelia, Apartado postal 3-72, 58090 Morelia, 
Michoacan, Mexico 
}

\date{Last updated ; in original form}
\pubyear{}

\begin{document}
\label{firstpage}
\pagerange{\pageref{firstpage}--\pageref{lastpage}}
\maketitle


\begin{abstract}

A significant fraction of planetary nebulae (PNe) exhibit collimated outflows, 
distinct narrow kinematical components with notable velocity shifts with 
respect to the main nebular shells typically associated with low-ionization 
compact knots and linear or precessing jet-like features.  
We present here a spatio-kinematical investigation of a sample of 
twelve PNe with morphologies in emission lines of low-ionization 
species suggestive of collimated outflows.  
Using archival narrow-band images and our own high-dispersion long-slit echelle 
spectra, we confirm the presence of collimated outflows in Hen\,2-429, J\,320, 
M\,1-66, M\,2-40, M\,3-1, and NGC\,6210 and possibly in NGC\,6741, for which 
the spatio-kinematical data can also be interpreted as a pair of bipolar lobes. 
The presence of collimated outflows is rejected in Hen\,2-47, 
Hen\,2-115, M\,1-26, and M\,1-37, but their morphology and 
kinematics are indicative of the action of supersonic
outflows that have not been able to pierce through the 
nebular envelope.  
In this sense, M\,1-66 appears to have experienced a similar interaction 
between the outflow and nebular envelope, but, as opposed to these four 
PNe, the outflow has been able to break through the nebular envelope.  
It is suggested that the PNe without collimated outflows in our 
sample are younger or descend from lower mass progenitors than 
those that exhibit unambiguous collimated outflows.

\end{abstract}


\section{Introduction}

Planetary nebulae (PNe) are the final stages of the life of low- and 
intermediate-mass stars (1--8 M$_\odot$), before they become white 
dwarfs. 
As such stars evolve through the asymptotic giant branch (AGB), 
they experience episodes of heavy mass-loss through a dense and 
slow AGB wind.  
Once the stellar core is exposed, these stars leave the AGB and their
effective temperatures increase while they develop fast and tenuous
stellar winds.  
The increasing UV stellar flux will ionize the slow 
AGB wind to give birth to a PN.

A large number of PNe exhibit fast collimated outflows, typically
described as distinct narrow kinematical components with notable
velocity shifts with respect to the main nebular shells.  
Examples abound in the literature: 
IC\,4634 \citep{GMR08}, 
Hu\,2-1 \citep{M95}, 
NGC\,7354 \citep{CVM10}, 
KjPn8 \citep{LVR95}, 
NGC\,6337 \citep{C00}, and 
M\,1-32 \citep{RGVP17}.
The reader is referred to our recent compilation and statistical 
investigation of collimated outflows in PNe \citep{GRO19}.

Collimated outflows are commonly associated with low-ionization compact knots 
and linear or precessing jet-like features \citep{GCM01}, but the sample of 
collimated outflows has diverse spatio-kinematical properties, which may point 
to different populations with different origins \citep{GRO19}.  
Since collimated outflows in PNe are thought to play notable
effects in their shaping and even formation \citep{ST98}, a
critical spatio-kinematical assessment of the nature of PN
features classified as possible collimated outflows based on
their morphologies and excitation conditions is most necessary.

In this paper, we present long-slit high-dispersion spectroscopic 
observations of a sample of PNe whose morphologies, as determined from 
narrow-band images, are indicative of the presence of collimated outflows.
The spatio-kinematical information derived from these observations
along the position angle (PA) of the possible collimated outflow
has been used to confirm (or reject) their nature and to derive 
their systemic radial velocities.
The sample of PNe and the observations are described in \S2,
the spatio-kinematical data of each PN is analysed in \S3, and
a discussion and a short summary are given in \S4 and \S5, 
respectively.

%
%

\begin{table*}
\centering
\caption*{Sample and Observations of PNe with Possible Collimated Outflows} 
\begin{tabular}{lllclccc}
\hline \hline 
\multicolumn{1}{l}{Source} & 
\multicolumn{3}{c}{\underline{~~~~~~~~~~~~~~~~~~~~~~~~~~~~~~Imagery~~~~~~~~~~~~~~~~~~~~~~~~~~~~~~}}       & 
\multicolumn{4}{c}{\underline{~~~~~~~~~~~~~~~~~~~~~~~~~~~~~~~~~~~~~~~~Spectroscopy~~~~~~~~~~~~~~~~~~~~~~~~~~~~~~~~~~~~~~~~}}  \\
\multicolumn{1}{l}{} & 
\multicolumn{1}{c}{Telescope \& Camera} & 
\multicolumn{1}{c}{Band} & 
\multicolumn{1}{c}{Resolution} & 
\multicolumn{1}{c}{Telescope \& Spectrograph} & 
\multicolumn{2}{c}{Resolution} & 
\multicolumn{1}{c}{PA} \\ 
\multicolumn{1}{l}{} & 
\multicolumn{1}{l}{} & 
\multicolumn{1}{l}{} & 
\multicolumn{1}{c}{($^{\prime\prime}$)} & 
\multicolumn{1}{l}{} & 
\multicolumn{1}{c}{(km~s$^{-1}$)} & 
\multicolumn{1}{c}{($^{\prime\prime}$)} & 
\multicolumn{1}{c}{($^\circ$)} \\

\hline	

Hen\,2-47  & \emph{HST}  WFPC2  & [N~{\sc ii}] & 0\farcs1 & Blanco Echelle & 8.0 & 1.4 &   0, 64 \\
Hen\,2-115 & \emph{HST}  WFPC2  & H$\alpha$    & 0\farcs1 & Blanco Echelle & 8.0 & 1.4 &   9, 275 \\
Hen\,2-429 & NOT         ALFOSC & [N~{\sc ii}] & 0\farcs7 & WHT    UES     & 6.5 & 1.2 &  90 \\
IC\,4776   & VLT         FORS2  & [O~{\sc ii}] & 1\farcs0 & Blanco Echelle & 8.0 & 1.7 &  45 \\
J\,320     & \emph{HST}  WFPC2  & H$\alpha$    & 0\farcs1 & Blanco Echelle & 8.0 & 1.8 & 306, 338, 348 \\
M\,1-26    & \emph{HST}  WFPC2  & H$\alpha$    & 0\farcs1 & Blanco Echelle & 8.0 & 1.5 &  38, 82, 325 \\
M\,1-37    & \emph{HST}  WFPC2  & H$\alpha$    & 0\farcs1 & Blanco Echelle & 8.0 & 1.3 &  39, 309 \\ 
M\,1-66    & \emph{HST}  WFPC2  & H$\alpha$    & 0\farcs1 & WHT    UES     & 6.5 & 2.2 & 311 \\
M\,2-40    & \emph{HST}  WFPC2  & H$\alpha$    & 0\farcs1 & WHT    UES     & 6.5 & 1.1 &  80 \\
M\,3-1     & \emph{HST}  WFPC2  & [N~{\sc ii}] & 0\farcs1 & Blanco Echelle & 8.0 & 1.8 &  72, 318, 358 \\
NGC\,6210  & \emph{HST}  WFPC2  & H$\alpha$    & 0\farcs1 & Blanco Echelle & 8.0 & 1.8 & 312, 336 \\
NGC\,6741  & \emph{HST}  WFPC2  & [N~{\sc ii}] & 0\farcs1 & Blanco Echelle & 8.0 & 1.3 &  18, 77, 302 \\

\hline
\end{tabular}
\label{tab:obs}
\end{table*}

\section{Observations}

\subsection{Imagery}

Collimated outflows in PNe are best identified in narrow-band
images of emission lines of low-ionization species of
[N~{\sc ii}], [O~{\sc i}], [O~{\sc ii}], or [S~{\sc ii}] 
\citep{GCM01}.  
We have selected a sample of PNe with narrow-band images
in [N~{\sc ii}], [O~{\sc ii}] or H$\alpha$ emission lines 
indicative of the presence of collimated outflows with no 
available kinematical data.
The sample is presented in Table~\ref{tab:obs}.
Most sources have high-quality narrow-band images obtained
with the Wide-Field Planetary Camera 2 (WFPC2) on board of
the \emph{Hubble Space Telescope} (\emph{HST}), except
IC\,4776, whose narrow-band image in the [O~{\sc ii}]
emission line was obtained with the instrument FORS2 at
the Very Large Telescope (VLT) of the European Southern
Observatory (ESO) in Chile \citep{Sowickar17}, and 
Hen\,2-429, whose narrow-band image in the [N~{\sc ii}] emission line 
was obtained with the instrument ALFOSC at the Nordic Optical Telescope 
(NOT) of the Observatorio de El Roque de los Muchachos (ORM) on the island
of La Palma in Spain \citep{MGS96}.

\subsection{Kinematical data}

Spatially-resolved high-dispersion echelle spectroscopic observations 
were obtained along a series of long-slit apertures placed across 
nebular features of interest.  
The Utretch Echelle Spectrograph (UES) on the 4.2m William Herschel Telescope 
(WHT) of the ORM and the echelle spectrograph on the Cerro Tololo Interamerican
Observatory (CTIO) 4m V\'\i ctor Blanco (aka Blanco) telescope were used for
these observations.


The high-dispersion spectroscopic observations of the northern sources 
were obtained on 13 July 1995, 6 August 1996, and 10 December 1998 using
UES at the WHT (Tab.~\ref{tab:obs}).  
The spectrograph was used in its long-slit mode with a narrow-band 
filter that isolates the echelle order in the spectral range from 6530 
to 6600 \AA, thus including the H$\alpha$ and [N~{\sc ii}] 
$\lambda\lambda$6548,6583 emission lines.  
A 79 line mm$^{-1}$ echelle grating was also used.  
The data were recorded with the  Tektronix CCD detector, providing 
a spatial scale of 0\farcs36 pixel$^{-1}$ and a sampling of 3.2 
km~s$^{-1}$~pixel$^{-1}$ along the dispersion direction.  
The slit has an unvignetted length of 160$^{\prime\prime}$ and its 
width was set to 1\farcs1, resulting in an instrumental resolution 
of 6.5 km~s$^{-1}$.
The angular resolution, determined by the seeing, varied between 
1\farcs1 and 2\farcs2 (Tab.~\ref{tab:obs}).  
Individual spectra have typical exposure times of 1,800 s.

The high-dispersion spectroscopic observations of the southern sources 
were obtained on 24-25 December 2001 and 22-25 June 2002 using the
echelle spectrograph on the CTIO Blanco telescope (Tab.~\ref{tab:obs}).  
The spectrograph was also used in its long-slit mode with a narrow-band 
$FWHM\sim50$ \AA\ filter to isolate the echelle order including the 
H$\alpha$ and [N~{\sc ii}] $\lambda\lambda$6548,6583 emission lines.  
The 79 line mm$^{-1}$ echelle grating and the long-focus red camera 
were used, resulting in a reciprocal dispersion of 3.4 \AA~mm$^{-1}$. 
The data were recorded with the SITe 2K CCD \#6, providing a spatial scale 
of 0\farcs26 pixel$^{-1}$ and a sampling of 3.7 km~s$^{-1}$~pixel$^{-1}$ 
along the dispersion direction.  
The slit has an unvignetted length of 3$^\prime$ and its width 
was set to 0\farcs9, resulting in an instrumental resolution of 
8.0 km~s$^{-1}$. 
The angular resolution, determined by the seeing, varied between 
1\farcs3 and 1\farcs8 (Tab.~\ref{tab:obs}).    
Individual spectra have typical exposure times of 1,800 s.

The spectra were reduced using standard IRAF tasks for 
two-dimensional spectra. 
The wavelength scale and geometrical distortion were corrected using a 
two-dimensional fit to arc exposures obtained using Th-Ar calibration 
lamps inmediately before or after the science exposure of each source.  
The deviation of the residuals of the two-dimensional 
fit to the Th-Ar arcs is found to be better than 0.004 
\AA\ (0.2~km~s$^{-1}$).  
The telluric emission, which includes the geocoronal H$\alpha$ 
line, was removed by fitting and subtracting the background 
using low-order polynoms.
The echelle observations were made with the slit oriented along 
different position angles and placed at various offsets from the 
central stars in order to sample different morphological features 
of interest (Tab.~\ref{tab:obs}).

\section{Results}

The narrow-band images and position-velocity (PV) maps of selected
emission lines along PAs of interest are presented in Figures 1 to
12 using a hyperbolic sine inverted gray-scale.  
The images are overlaid with the location of the slits used
to obtain kinematical information (Tab.~\ref{tab:obs}).
The brightest regions in the PV maps are shown 
saturated and overlaid with contours to display 
simultaneously information of the faintest and 
brightest features of these PV maps.
These PV maps have been corrected to the Local Standard of Rest (LSR)
velocity system and the spatial offsets have been measured from the
central star or from its projection onto the slit.

The radial velocity in the LSR ($v_{\rm LSR}^{\rm sys}$) and heliocentric 
($v_{\rm hel}^{\rm sys}$) systems and nebular expansion velocity of these
sources ($v_{exp}$) are listed in columns 3 to 5 of Table~\ref{tab:result}, 
respectively.  
The radial velocities have been derived from the H$\alpha$ emission line 
profiles of the main nebular shells at the location of the central star, 
whereas the profiles of the narrower [N~{\sc ii}] $\lambda$6583 \AA\ 
emission line have been used to determine the nebular expansion velocities. 
When the emission line profile is resolved into receding 
(red-shifted) and approaching (blue-shifted) components, the 
systemic radial velocity is computed as the average between the 
centroid of both components and the nebular expansion velocity 
as their semi-difference.
Typically, the velocity from an emission line can be determined with an
accuracy $\sim$10\% the spectral resolution, which combined with the
wavelength calibration uncertainty of 0.2 km~s$^{-1}$, implies velocity
uncertainties within $\pm$1.0 km~s$^{-1}$.  
In cases when the emission line profile is unresolved, the systemic radial 
velocity is computed as the centroid of the emission line profile and the 
nebular expansion velocity as the half width of the line profile once the 
instrumental and thermal widths have been substracted quadratically.  
The radial velocities of sources in Table~\ref{tab:result} 
are generally in agreement ($\leq$10 km~s$^{-1}$) with previous 
available measurements, as for 
Hen\,2-115 \citep[$v_{\rm hel} \simeq -56.2$ km~s$^{-1}$,][]{DAZ98}, 
Hen\,2-249 \citep[$v_{\rm hel} \simeq 30$ km~s$^{-1}$,][]{GVL99}, 
IC\,4776 \citep[$v_{\rm hel} \simeq 16.3$ km~s$^{-1}$ and 
$v_{\rm LSR} \simeq 27.9$ km~s$^{-1}$,][]{DAZ98,MD92}, 
J\,320 \citep[$v_{\rm hel} \simeq -25$ km~s$^{-1}$,][]{HBL04}, 
M\,1-26 \citep[$v_{\rm hel} \simeq -22$ km~s$^{-1}$,][]{OS85}, 
M\,1-37 \citep[$v_{\rm hel} \simeq 213$ km~s$^{-1}$,][]{RSL17}, 
M\,1-66 \citep[$v_{\rm hel} \simeq 27.9$ km~s$^{-1}$,][]{DAZ98}, 
M\,2-40 \citep[$v_{\rm hel} \simeq 90$ km~s$^{-1}$,][]{BDF92}, 
M\,3-1 \citep[$v_{\rm hel} \simeq 69.5$ km~s$^{-1}$,][]{ST83}, 
and 
NGC\,6741 \citep[$v_{\rm hel} \simeq 41.3$ km~s$^{-1}$,][]{ST83}.  
Only in Hen\,2-47 the difference between the radial velocity 
in Table~\ref{tab:result} and the heliocentric radial velocity 
of $-17.7$ km~s$^{-1}$ reported by \citet{DAZ98} is larger than 
10 km~s$^{-1}$.

The spatio-kinematical properties of their collimated outflows, 
including their identification, morphology, radial velocities 
with respect to the systemic velocity ($v_{\rm r}^{\rm outflow}$), 
and projected linear sizes ($\delta r$) are listed in columns 
6 to 9 of Table~\ref{tab:result}, respectively.
Hereafter, we will refer to the radial velocity of the outflow with
respect to the nebular systemic velocity as the systemic velocity.  
In most cases, kinematical information was available for both the approaching
and receding outflow components, and thus the systemic velocity was derived as 
the semi-difference between the radial velocities of these two components.

\begin{table*}
\centering
\caption*{Spatio-kinematical Information of the Sample of PNe with Possible Collimated Outflows}
\begin{tabular}{lrrrrclrr}
\hline \hline 
Source     & 
\multicolumn{1}{c}{Distance$^\star$} & 
\multicolumn{1}{c}{$v_{\rm LSR}^{\rm sys}$} & 
\multicolumn{1}{c}{$v_{\rm hel}^{\rm sys}$} & 
\multicolumn{1}{c}{$v_{exp}$} & 
Feature ID & Morphology & 
\multicolumn{1}{c}{$v_{\rm r}^{\rm outflow}$}  & 
\multicolumn{1}{c}{$\delta r$} \\
           &
\multicolumn{1}{c}{(kpc)} & 
\multicolumn{1}{c}{(km~s$^{-1}$)} & 
\multicolumn{1}{c}{(km~s$^{-1}$)} & 
\multicolumn{1}{c}{(km~s$^{-1}$)} & 
  	   &
           &   
\multicolumn{1}{c}{(km~s$^{-1}$)} & 
\multicolumn{1}{c}{(pc)} \\
\hline

Hen\,2-47  &  3.8~~~~ &  $-$16.3 &   $-$4.0 &   11.0~~~  &     N2-S3        & V-shape            & 23.5~~~~~~ & 0.074 \\
           &          &          &          &            &     N4-S2        & V-shape            & 22.9~~~~~~ & 0.071 \\
Hen\,2-115 &  5.0~~~~ &  $-$64.8 &  $-$63.6 &   13.4~~~  & PA=95$^{\circ}$~~ & V-shape            &  5.9~~~~~~ & 0.069 \\
Hen\,2-429 &  3.7~~~~ &  $+$45.2 &  $+$26.7 &   30.6~~~  & PA=90$^{\circ}$~~ & precessing outflow &  5.2~~~~~~ & 0.135 \\
IC\,4776   &  4.4~~~~ &  $+$23.0 &  $+$14.3 & $\dots$~~~ &      A-B         & bipolar lobes      & 49.6~~~~~~ & 0.100 \\ 
           &          &          &          &            &      C-D         & outflow?           & 84.1~~~~~~ & 0.174 \\     
J\,320     &  5.8~~~~ &  $-$44.4 &  $-$29.8 &    16.0~~~ & PA=348$^{\circ}$  & outflow            & 19.9~~~~~~ & 0.276 \\
           &          &          &          &            & PA=338$^{\circ}$  & outflow            & 34.0~~~~~~ & 0.219 \\
           &          &          &          &            & PA=306$^{\circ}$  & outflow            & 24.8~~~~~~ & 0.104 \\
M\,1-26    &  2.1~~~~ &  $-$14.4 &  $-$21.4 &  $<$7.0~~~ & PA=82$^{\circ}$~~ & bow-shock          & 38.2~~~~~~ & 0.038 \\
           &          &          &          &            & PA=145$^{\circ}$  & bow-shock          & 47.7~~~~~~ & 0.035 \\
M\,1-37    & 14.4~~~~ & $+$229.2 & $+$218.7 &  11.0~~~   & PA=129$^{\circ}$  & V-shape            &  7.6~~~~~~ & 0.142 \\
M\,1-66    &  6.4~~~~ &  $+$39.2 &  $+$22.6 &  20.0~~~   & PA=131$^{\circ}$  & outflow            &  7.0~~~~~~ & 0.106 \\
M\,2-40    &  5.4~~~~ &  $+$98.9 &  $+$82.8 &  17.6~~~   & PA=88$^{\circ}$~~ & outflow            & 16.0~~~~~~ & 0.123 \\
M\,3-1     &  4.5~~~~ &  $+$46.3 &  $+$65.6 &  24.5~~~   & PA=318$^{\circ}$  & precessing outflow & 14.9~~~~~~ & 0.268 \\
           &          &          &          &            & PA=358$^{\circ}$  & precessing outflow & 13.2~~~~~~ & 0.168 \\
NGC\,6210  &  2.1~~~~ &  $-$26.7 &  $-$45.7 & 34.2~~~    &       A          & precessing outflow & 19.5~~~~~~ & 0.084 \\
           &          &          &          &            &       B          & outflow            & 30.6~~~~~~ & 0.045 \\
           &          &          &          &            &       C          & precessing outflow & 29.4~~~~~~ & 0.174 \\
NGC\,6741  &  3.2~~~~ &  $+$56.9 &  $+$40.2 & 23.4~~~    &      A-B         & outflow            & 22.7~~~~~~ & 0.126 \\
           &          &          &          &            &       C          & outflow            &  7.3~~~~~~ & 0.100 \\

\hline
\multicolumn{9}{l}{$\star$ Distances adopted from \citet{FPB16}} \\
\end{tabular}
\label{tab:result}
\end{table*}

\begin{figure*}
\includegraphics[width=2.0\columnwidth]{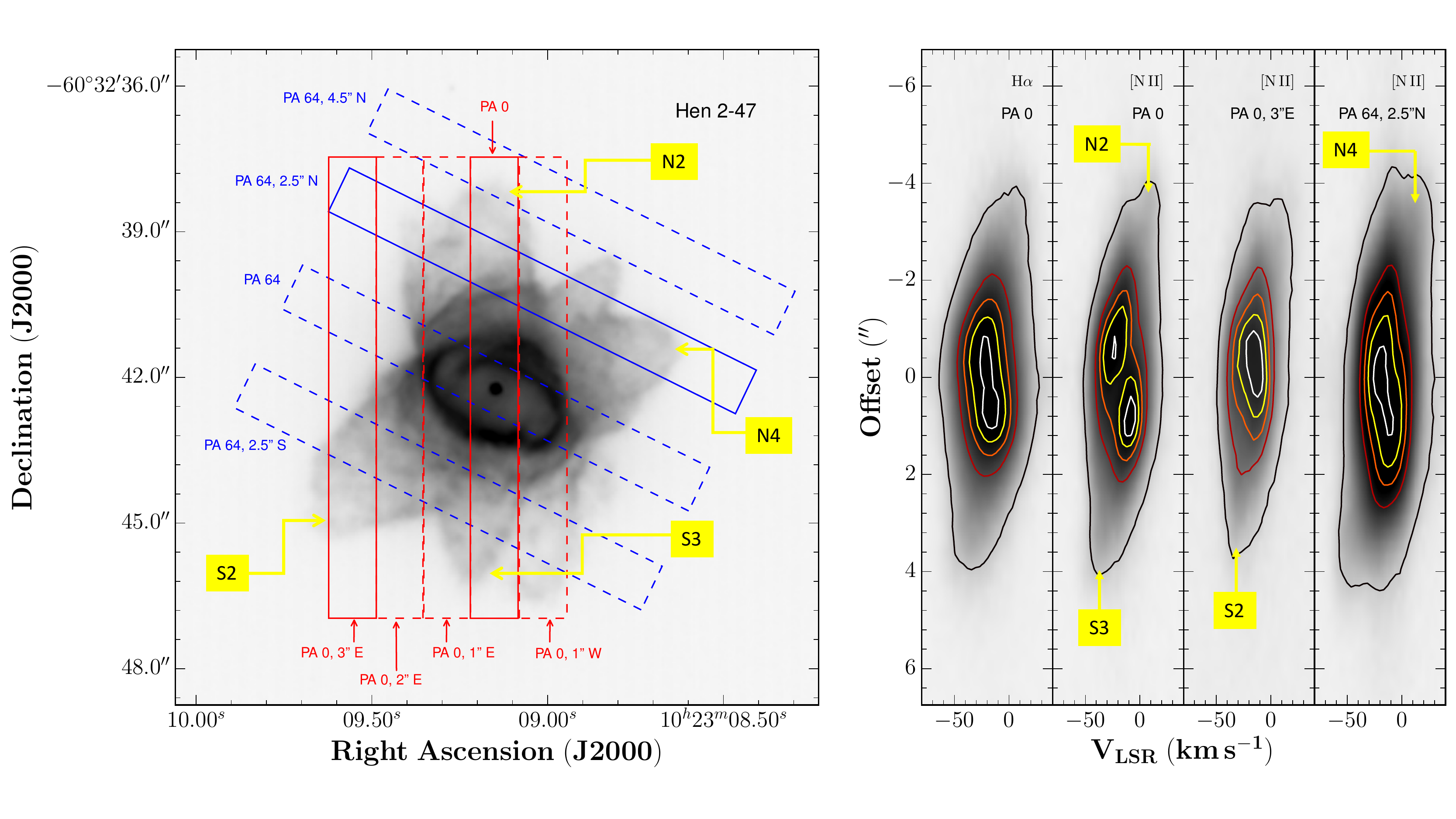}
\caption{
\emph{HST} WFPC2 F658N [N~{\sc ii}] image (left) and 
H$\alpha$ and [N~{\sc ii}] $\lambda$6583 PV maps derived
from CTIO 4m echelle spectra (right) of Hen\,2-47.
The names of the different morphological feature follow \citet{Sahai00}.
The positions of the slits are overlaid on the image
on the left panel;  the slits whose PV maps are shown
on the right panels are drawn with solid lines, those
not shown are drawn with dashed lines.  
}
\label{hen2-47}
\end{figure*}

\subsection{Hen\,2-47}

Hen\,2-47 (PN\,G285.6$-$02.7) is a young PN with a highly asymmetric multipolar 
morphology consisting of four pairs of V-shaped lobes and 
two rings with point-symmetric brightness distribution 
\citep{Sahai00}.
Its shape is very similar to that of M\,1-37 (see \S\ref{m137}) 
and both of them are consequently nicknamed the \textit{Starfish 
Twins}.
The complex shape of these young PNe has been proposed to result from
the interaction of high-velocity collimated outflows with the nebular 
envelope in the late AGB or at the beginning of the post-AGB phase 
\citep{ST98}.

The kinematics of Hen\,2-47 is investigated here using 9 slit positions
oriented along PA = 0$^\circ$ and 64$^\circ$ with different offsets from
the central star of the PN (CSPN), as listed in Table~1.
The most relevant [N~{\sc ii}] $\lambda$6583 PV maps are presented in
Figure~\ref{hen2-47}.

The expansion velocity of Hen\,2-47, as derived from the
emission profile of the [N~{\sc ii}] line in the slit
across the central star along PA=64$^\circ$ (not shown in
Figure~\ref{hen2-47}, is found to be 11.0 km~s$^{-1}$.

\begin{figure*}
\includegraphics[width=2.0\columnwidth]{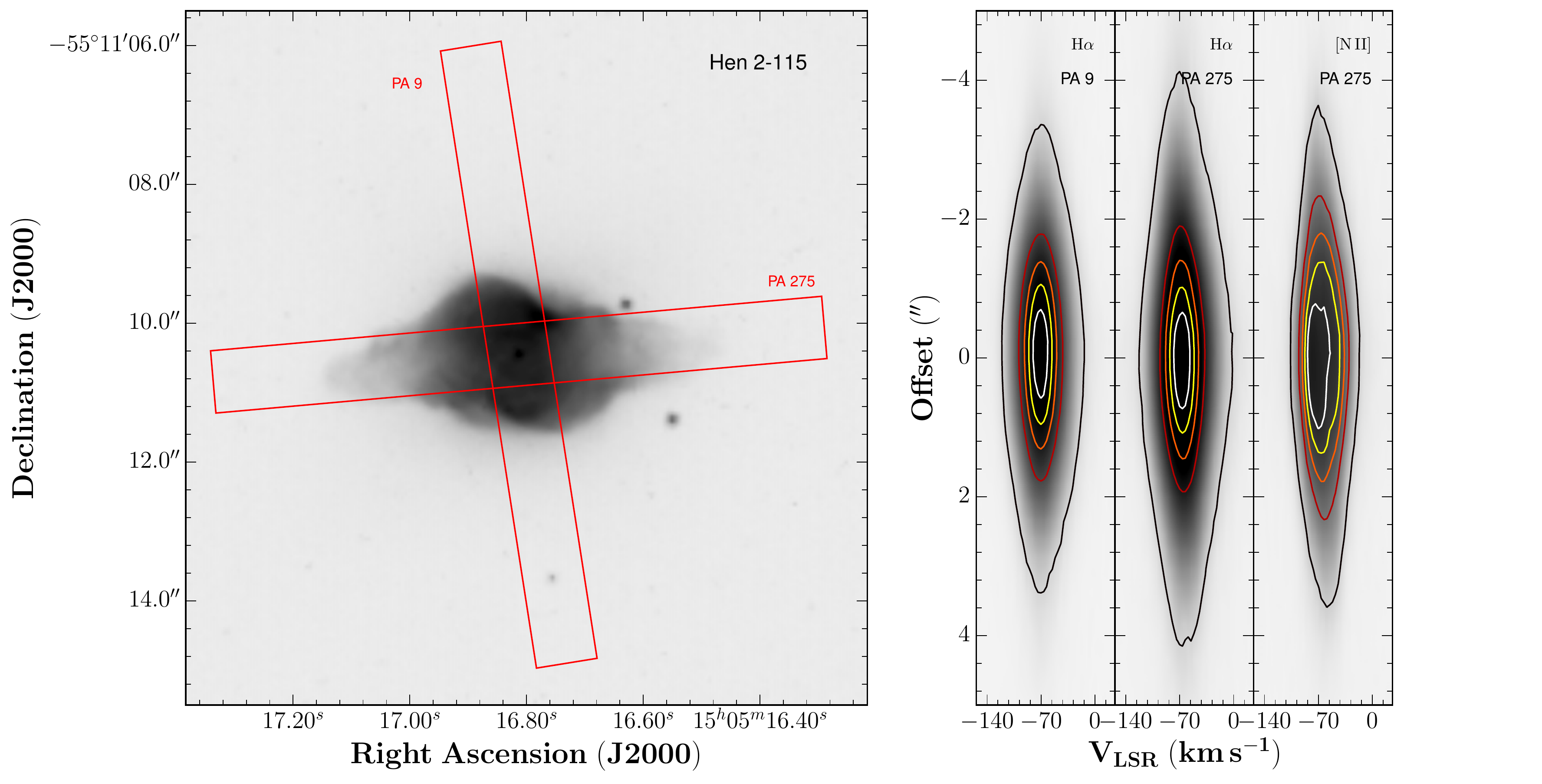}
\caption{
Same as Figure~\ref{hen2-47} for Hen\,2-115.  
In this case, the \emph{HST} image was obtained using the F656N H$\alpha$ 
filter.  
}
\label{hen2-115}
\end{figure*}

Following \citet{Sahai00}'s naming convention, features N2 and S3
are registered by the slit at PA = 0$^\circ$ across the CSPN, feature
S2 is registered by the slit at PA = 0$^\circ$ and 3$^{\prime\prime}$
East of the CSPN, and its counterpart N4 is registered by the slit
at PA = 64$^\circ$ offset by 2\farcs5 North from the CSPN.  
The PV maps indeed reveal that these features have 
kinematical properties differing from those of the 
main nebula.  
Multiple velocity components are detected wherever the slit 
intersects several V-shaped lobes, indicating that these 
have different radial velocities.  
From our data, we have been able to infer systemic
velocities $v_{\rm r}^{\rm outflow}$ of $\pm$23.5
km~s$^{-1}$ and $\pm$22.9 km~s$^{-1}$ for the pairs
of features N2-S3 and N4-S2, respectively.
Kinematically, these outflows do not look very collimated, 
in agreement with the impression derived from the H$\alpha$ 
image. 
These features have projected distances from the CSPN of $2.3\times 10^{17}$ 
cm (0.074 pc) and $2.2\times 10^{17}$ cm (0.071 pc), respectively.

\subsection{Hen\,2-115}

Hen\,2-115 (PN\,G321.3$+$02.8) is classified as an elliptical nebula
\citep{SCS93b}, although the \emph{HST} WFPC2 H$\alpha$ image in
Figure~\ref{hen2-115} reveals a complex morphology consisting of a
bright bipolar structure around the central star and a pair of
narrow V-shaped lobes protruding along PA = 95$^{\circ}$ \citep{ST98},
which seem indicative of the effects of a bipolar outflow.  
The main nebula is surrounded by a diffuse halo \citep{SMV11}.

The kinematics of Hen\,2-115 is investigated using two slits oriented
along its minor axis at PA~9$^\circ$ and along its major axis and
collimated outflows at PA~95$^\circ$.  
The [N~{\sc ii}] PV map of the first slit (not shown in 
Figure~\ref{hen2-115}) has been used to derive a nebular 
expansion velocity $\simeq$13.4 km~s$^{-1}$.

The [N~{\sc ii}] $\lambda$6584 PV map along PA = 95$^\circ$ shown in 
Figure~\ref{hen2-115} reveals the presence of a low-velocity 
($v_{\rm r}^{\rm outflow} = \pm 5.9$ km~s$^{-1}$) component at the location of the 
V-shaped lobes.  
As for Hen\,2-47, the outflow does not seem particularly 
collimated and the data presented here fail to detect 
emission from a compact knot expanding ballistically at 
the tip of the V-shaped lobe, e.g.\ as in CRL\,618 
\citep{Balick_etal2013,Balick_etal2014}.  
This feature has a projected distance from the 
CSPN $\simeq$2$\times10^{17}$ cm (0.069 pc).

\subsection{Hen\,2-429}

\begin{figure*}
\includegraphics[width=2.0\columnwidth]{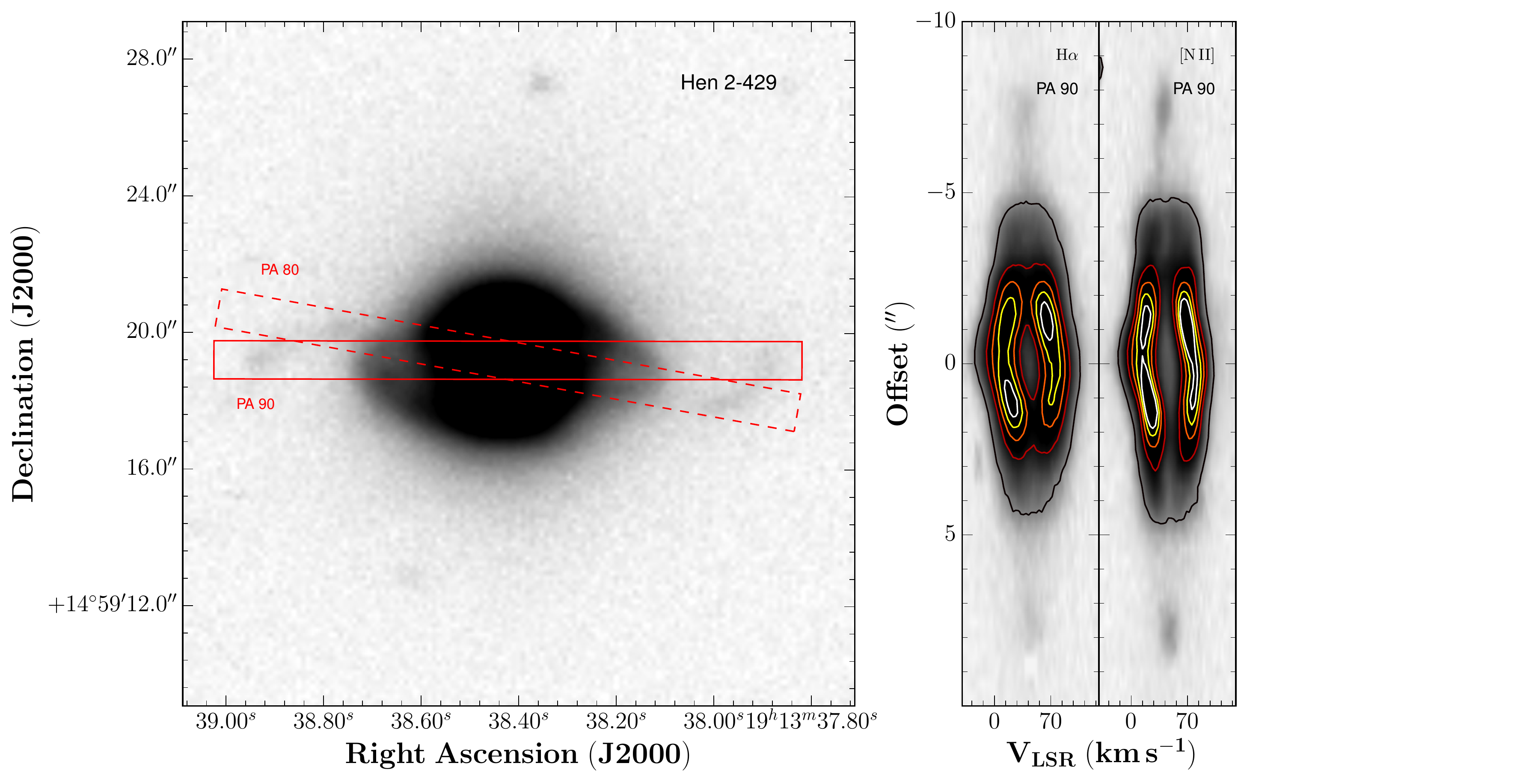}
\caption{
NOT ALFOSC [N~{\sc ii}] image (left) and H$\alpha$ and [N~{\sc ii}] 
$\lambda$6583 PV maps derived from WHT UES echelle spectra (right) 
of Hen\,2-429.  
Solid and dashed slits plotted on the image
as in Figure~\ref{hen2-47}.  
}
\label{hen2-429}
\end{figure*}

The [N~{\sc ii}] image of Hen\,2-429 (PN\,G048.7$+$01.9) in
Figure~\ref{hen2-429} shows an elliptical shell and a pair
of outer low-ionization point-symmetric filaments \citep{GVL99,GCM01}.

\begin{figure*}

\includegraphics[width=2.0\columnwidth]{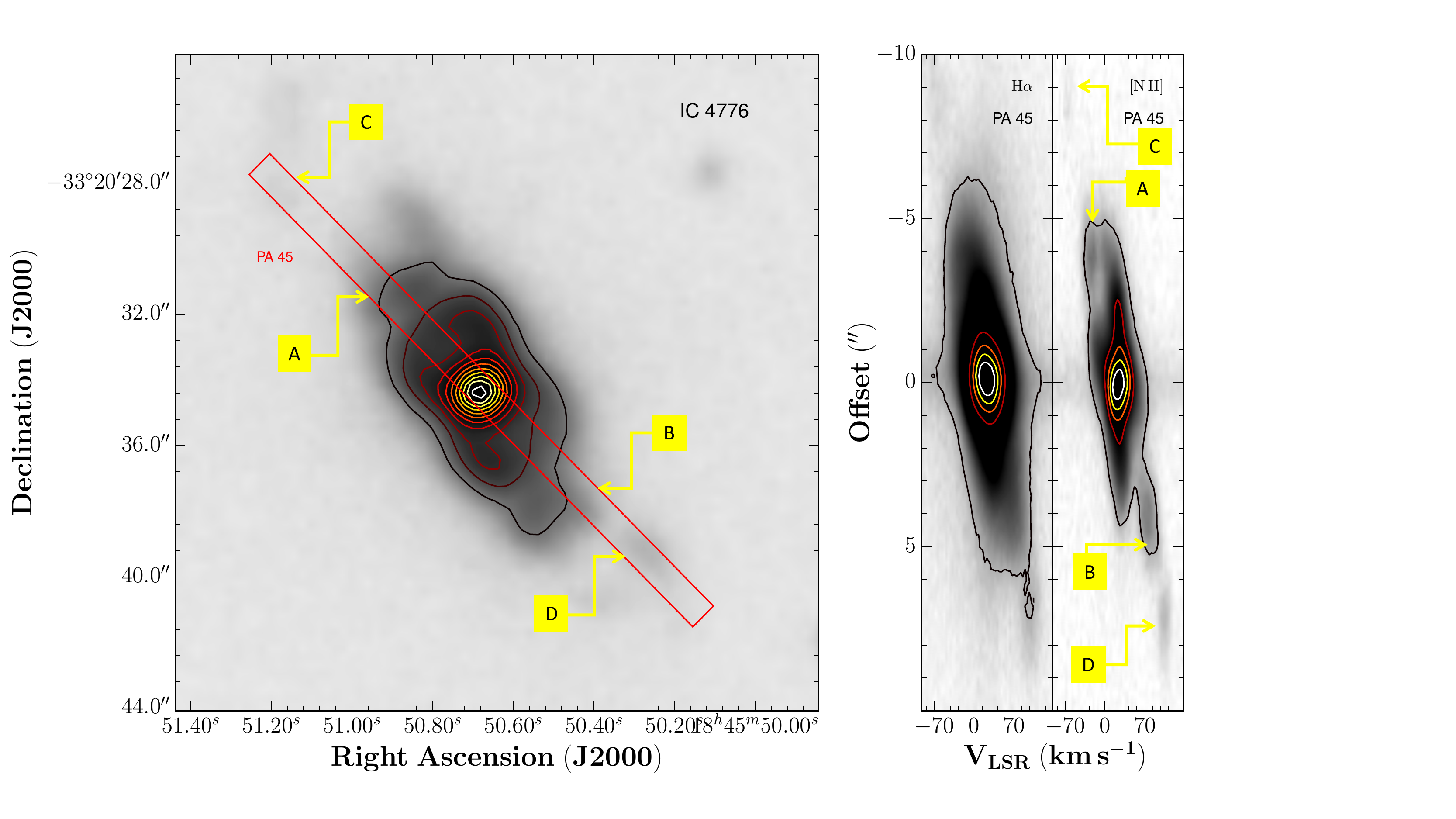}
\caption{
VLT FORS2 [O~{\sc ii}] image (left) and H$\alpha$ and [N~{\sc ii}] 
$\lambda$6583 PV maps derived from CTIO 4m echelle spectra (right) 
of IC\,4776.  
}
\label{ic4776}
\end{figure*}

Our kinematical investigation uses two slits placed at PA = 80$^\circ$ and 
90$^\circ$ across the central star.  
Since the point-symmetric outer filament is covered by both slits, 
only the [N~{\sc ii}] $\lambda$6583 PV map of the slit at PA = 90$^\circ$, 
registering emission at the tip of the filament, is shown in 
Figure~\ref{hen2-429}.  
The nebular expansion of Hen\,2-429 is well resolved in these 
PV maps.  
Its expansion velocity is found to be $\simeq$30.6 km~s$^{-1}$, 
in good agreement with that of 30 km~s$^{-1}$ reported by \citet{GVL99}.

The point-symmetric filaments are detected both in the H$\alpha$ 
and [N~{\sc ii}] PV maps as relatively bright and compact components 
that are moving at a systemic velocity $\pm$5.2 km~s$^{-1}$.  
The kinematical properties of these filaments was already discussed by 
\citet{GVL99}, who noted their low systemic velocity.  
Interestingly, there is faint emission between these 
knots and the main nebula, suggesting these are the 
tips of bow-shock-like features.  
Their projected distance from the CSPN is $4.2\times 10^{17}$ cm 
(0.14 pc).

\subsection{IC\,4776}

Recent VLT FORS2 narrow-band imaging of IC\,4776 (PN\,G002.0$-$13.4) in
the H$\alpha$+[N~{\sc ii}], [O~{\sc ii}], and [O~{\sc iii}] emission
lines have revealed an hourglass-shaped main nebula with its major axis
aligned along PA = 45$^{\circ}$ and a pair of precessing jet-like features
\citep{Sowickar17}.

Its kinematics is investigated here using a slit oriented along its
major axis (Figure~\ref{ic4776}).
The expansion velocity of the inner brightest nebular regions is 
lower than 8 km~s$^{-1}$, in agreement with previous estimates 
\citep[$\simeq 10.2$ km~s$^{-1}$,][]{Bianchi92}.

\begin{figure*}
\includegraphics[width=2.0\columnwidth]{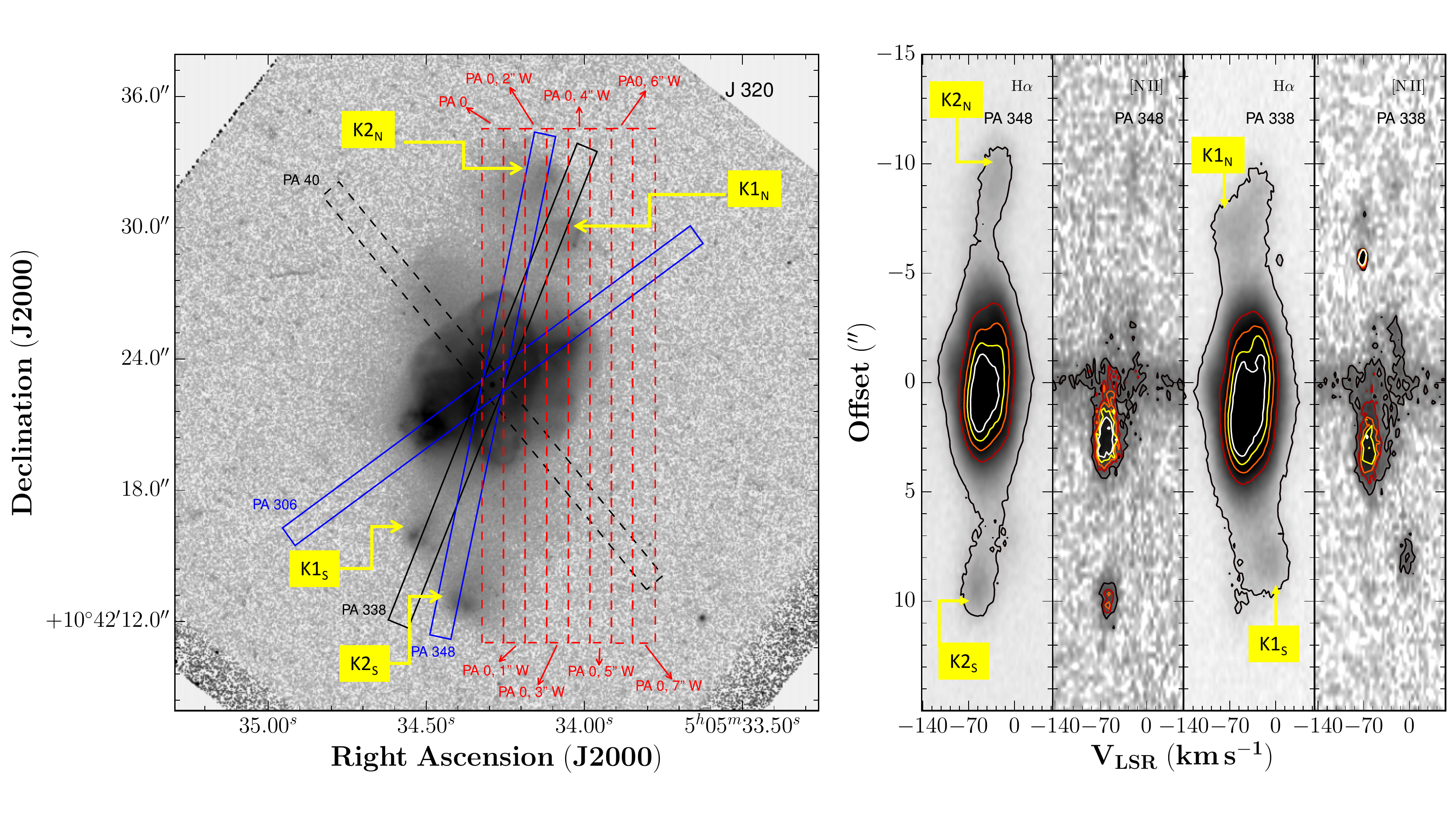}
\includegraphics[width=2.0\columnwidth]{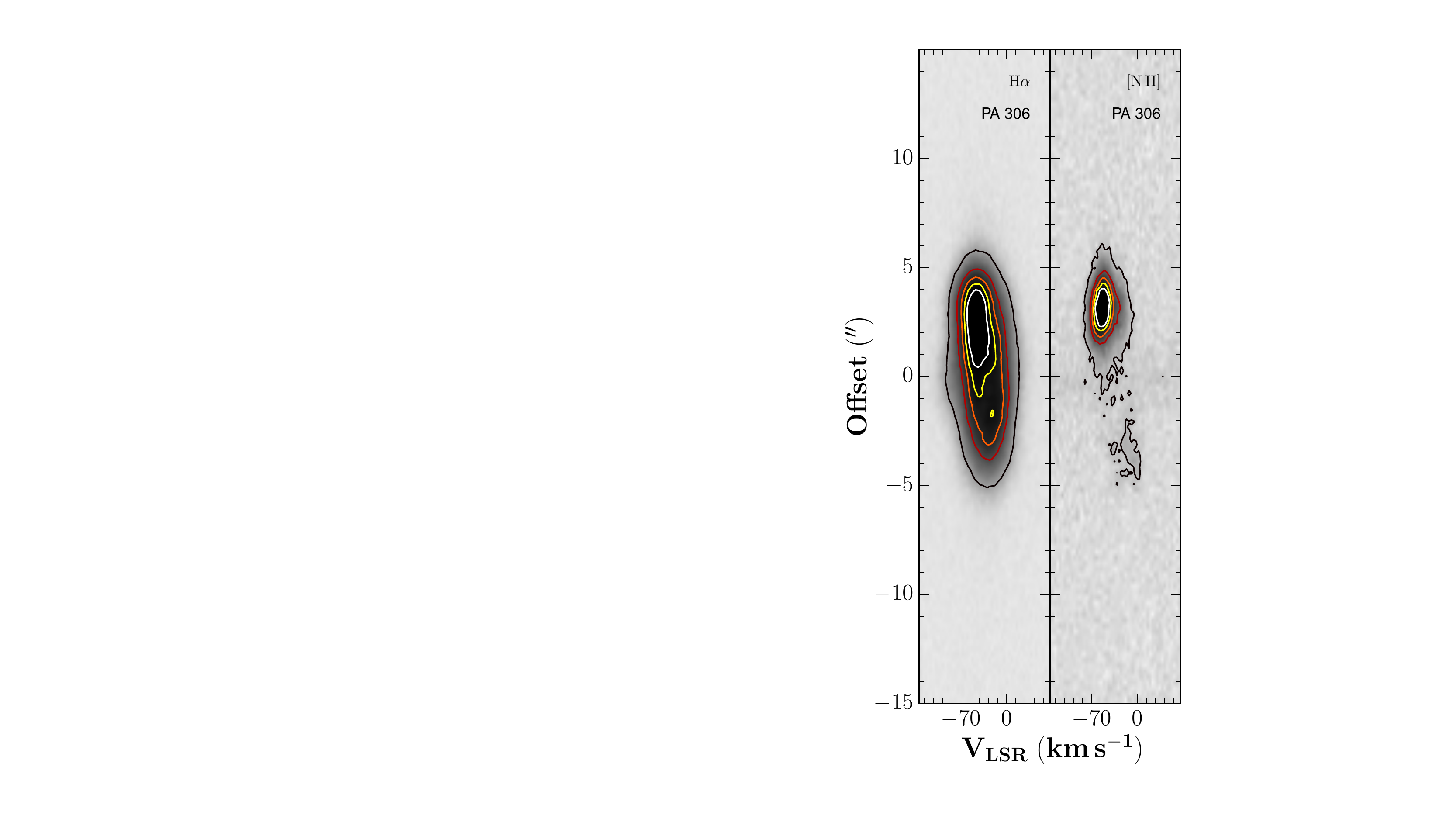}
\caption{
Same as Figure~\ref{hen2-47} for J\,320.
In this case, the \emph{HST} image was obtained using the F656N H$\alpha$
filter.
}
\label{j320}
\end{figure*}

The H$\alpha$ and [N~{\sc ii}] $\lambda$6583 PV maps reveal distinct
structural components for the main nebula:
an ellipsoidal shell in the H$\alpha$ line, but
an X-shaped structure in the [N~{\sc ii}] emission
line.  
Four features named as A, B, C, and D are marked in Figure~\ref{ic4776} 
as possible collimated outflows.  
The X-shaped pattern is revealing of a bipolar PN whose 
symmetry axis is tilted with the line of sight.  
This interpretation is basically consistent with that provided 
by \citet{Sowickar17}, but the extent of the bipolar lobes of 
$\simeq$12$^{\prime\prime}$ in size derived from our PV maps is 
larger than that of $\simeq$7$^{\prime\prime}$ derived from the 
simulated PV map presented in figure~4 by \citet{Sowickar17}
because we interpret that the A and B features in our PV maps 
close the X-shaped pattern onto bipolar lobes.  
The systemic velocity of features A and B is
found to be $\pm$52.2 km~s$^{-1}$.  
As for the outermost C and D pair of features, their systemic velocities
are larger, $\pm$84 km~s$^{-1}$, and they look like compact knots, although
we cannot discard they are the brightest regions of one additional pair of
point-symmetric bipolar lobes oriented along the same nebular axis 
\citep[e.g., M\,1-16 and M\,2-9][]{Schwarz1992,Schwarz_etal1997,Clyne_etal2015}.
These features have projected distances from the CSPN of 
$3.1\times 10^{17}$ cm (0.10 pc) and $5.4\times 10^{17}$ 
cm (0.17 pc), respectively.

\subsection{J\,320}

J\,320 (PN\,G190.3$-$17.7) is a spatially complex PN discovered by
\citet{Jonck16} and classified morphologically as point-symmetric
\citep{GCM01}.
It presents a bright elliptical inner rim, two bipolar lobes and several
pairs of jet-like features oriented along different directions \citep{CM17}.
The latter suggest that the ejecting direction is rotating, 
possibly with a time dependent velocity \citep{Lopez97}.

\begin{figure*}
\includegraphics[width=2.0\columnwidth]{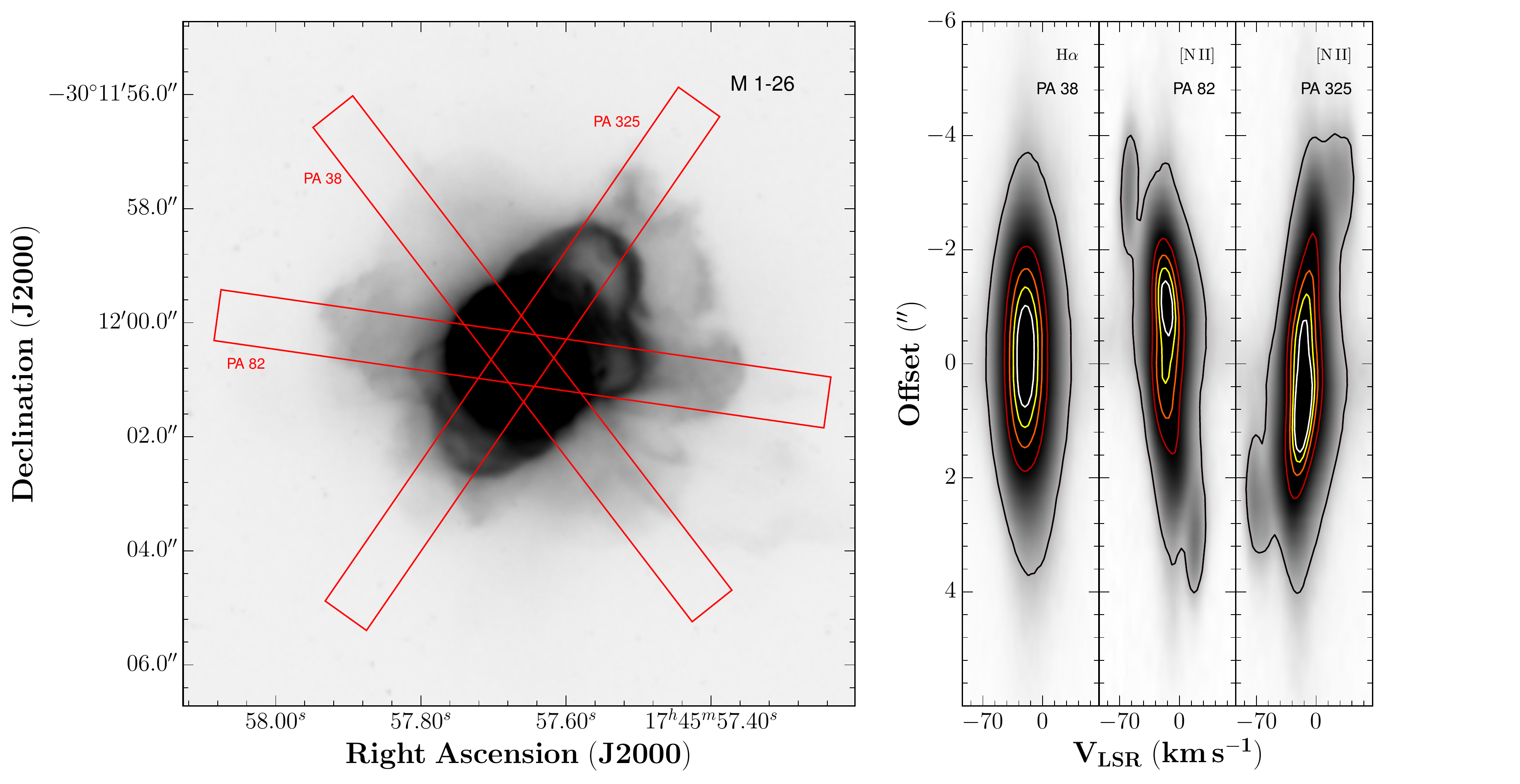}
\caption{
Same as Figure~\ref{hen2-47} for M\,1-26.  
In this case, the \emph{HST} image was obtained using the F656N H$\alpha$ 
filter.  
}
\label{m1-26}
\end{figure*}

To investigate the kinematics of J\,320, spectra were obtained using eight
parallel slits oriented North-South and with offsets of 1$^{\prime\prime}$
between positions (red dash boxes in Fig.~\ref{j320}-left) and four slits
oriented at PA = 40$^\circ$, 306$^\circ$, 338$^\circ$ and 348$^\circ$.
The nebula has an expansion velocity $\simeq 16$ km~s$^{-1}$, 
derived in this case from the H$\alpha$ PV maps across the CSPN.

The most interesting PV maps correspond to the slits at PA = 338$^\circ$
and 348$^\circ$ (Figure~\ref{j320}), which reveal that the bipolar lobes
tips are bright pairs of knots.
Following \citet{HBL04}'s naming convention for these knots, we mark
them in both the H$\alpha$ image and PV maps in Figure~\ref{j320} as
K1$_{\rm N}$ and K1$_{\rm S}$ for the knots detected in the slit at PA =
338$^\circ$, and K2$_{\rm N}$ and K2$_{\rm S}$ for the slit at PA = 348$^\circ$. 
The systemic velocity of these pairs of knots is 
$\pm$34.1 km~s$^{-1}$ for K1 and $\pm$19.9 km~s$^{-1}$ 
for K2, in agreement with the velocities reported by 
\citet{HBL04}.  
The slit at PA = 306$^\circ$ reveals in the main body a pair of knots, one
brighter than the other, with systemic velocity $\pm$24.8 km~s$^{-1}$. 
The projected distances from the CSPN of these features are 
$6.7\times 10^{17}$ cm (0.217 pc) for the pair of knots K1,
$8.5\times 10^{17}$ cm (0.275 pc) for the pair of knots K2, and
$3.2\times 10^{17}$ cm (0.105 pc) for the pair of knots along PA = 306$^\circ$.

\subsection{M 1-26}

Whereas ground-based images of M\,1-26 (PN\,G358.9$-$00.7) show an apparently
featureless elliptical disk \citep{GST97}, the \emph{HST} WFPC2 H$\alpha$
image in Figure~\ref{m1-26}-\emph{left} discloses it as a bright round shell 
surrounded by multiples loops and arcs \citep{ST98}. 
The multipolar morphology of M\,1-26, which makes it look 
like a \textit{rose}, prompted \citet{BS17} to include it 
among the sample of PNe likely shaped by triple stellar 
progenitors.

The kinematics of M\,1-26 has been investigated using three long-slit 
apertures placed along the minor nebular axis at the PA = 38$^\circ$, 
and at the location of outflow-like features along PA = 82$^\circ$ and 
145$^\circ$.  
The H$\alpha$ PV map for PA = 38$^\circ$ and [N~{\sc ii}] $\lambda$6583 
PV maps of the slit at PA = 82$^\circ$ and  PA = 145$^\circ$ are shown in 
Figure~\ref{m1-26}.
The expansion velocity of the main nebular shell is not resolved.  
We estimate an upper limit of 7 km~s$^{-1}$ based on the 
line width.

The outflow-like features registered by our long-slit observations indeed 
reveal systemic velocities of $\pm 38.2$ km~s$^{-1}$ at PA = 82$^\circ$ and
$\pm 47.7$ km~s$^{-1}$ at PA = 145$^\circ$, which are faster than the
expansion velocity of the nebular shell.
The velocities here detected certainly imply that the arcs 
and loops seen in optical images have suffered a notable 
acceleration, most likely produced by the action of collimated 
outflows whose tips are not clearly identified in the PV maps.  
The outflows have projected distances from the CSPN of 
$1.18\times 10^{17}$ cm (0.04 pc) and 
$1.07\times 10^{17}$ cm (0.03 pc), respectively.

 \begin{figure*}
\includegraphics[width=2.0\columnwidth]{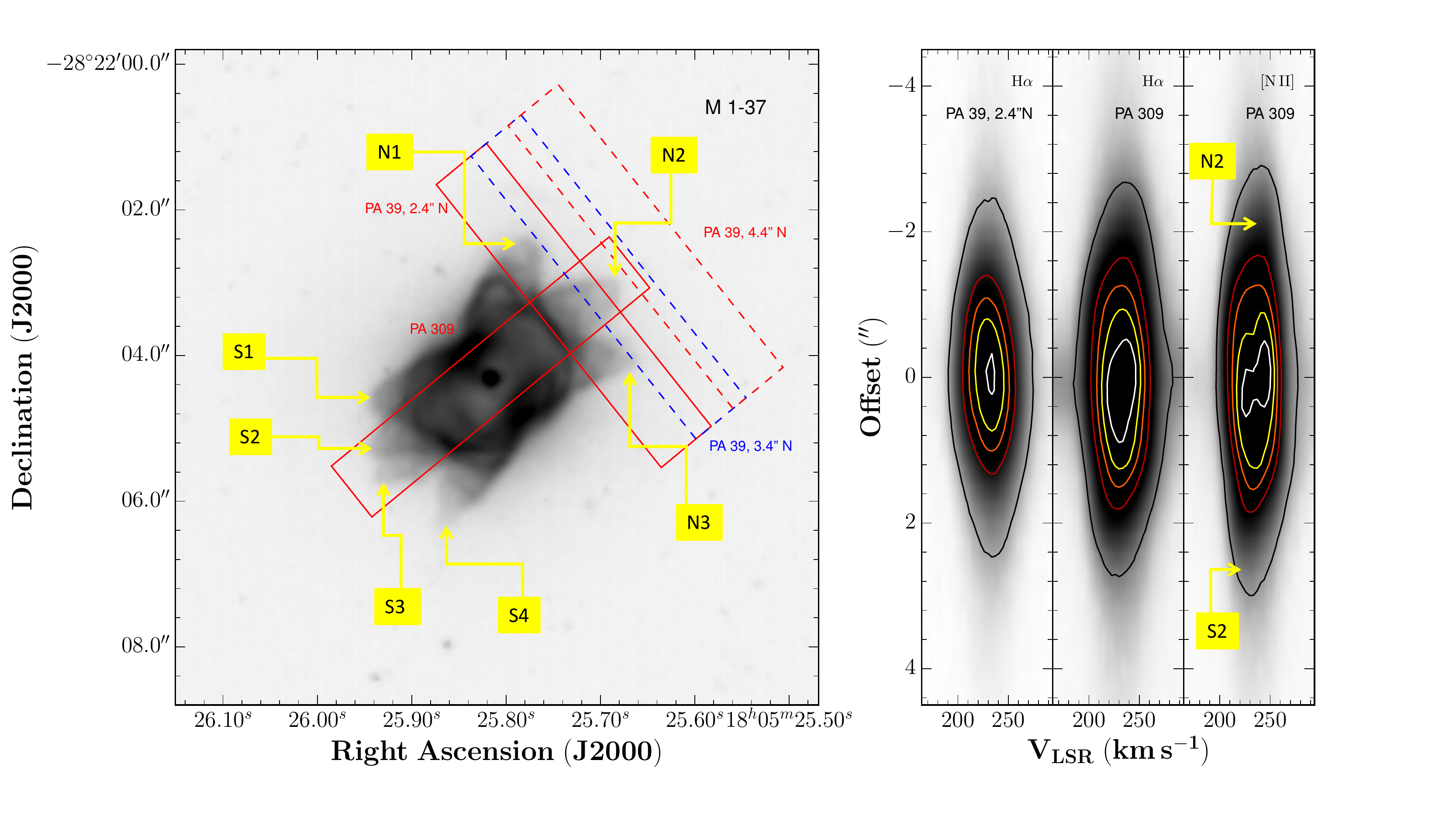}
\caption{
Same as Figure~\ref{hen2-47} for M\,1-37.  
In this case, the \emph{HST} image was obtained using the F656N H$\alpha$ 
filter.  
}
\label{m1-37}
\end{figure*}

\subsection{M 1-37}
\label{m137}

M\,1-37 (PN\,G002.6$-$03.4) is the \textit{Starfish Twin}
of Hen\,2-47 \citep{Sahai00}.
Its morphology, as shown by its H$\alpha$ image (Figure~\ref{m1-37}), 
is multipolar, with seven V-shaped lobes arranged along different 
directions from the highly distorted bipolar-shaped inner shell.  
Following \citet{Sahai00}'s naming convention for these lobes, 
we mark them from N1 to N3 at the North and S1 to S4 at the 
South in the H$\alpha$ image in Figure~\ref{m1-37}.  
The main nebula is surrounded by a faint halo.

The kinematics of M\,1-37 is investigated using a slit placed along its 
major axis (PA=129$^\circ$) and another one along PA = 39$^\circ$ offset from 
the central star to register the N1, N2, and N3 features (see the right panel 
of Fig.~\ref{m1-37}).  
A nebular expansion velocity of 11 km~s$^{-1}$ is estimated
from the PV maps of the slit at PA = 129$^\circ$.

As for its twin, Hen\,2-47, the V-shaped lobes do not show extremely 
narrow kinematical components, with a small systemic velocity for the
pair N2-S2 of $v_{\rm r}^{\rm outflow} = \pm 7.6$ km~s$^{-1}$.
These features have projected distances from the CSPN of 
4.4$\times$10$^{17}$ cm (0.14 pc).

\subsection{M 1-66}

\begin{figure*}
\includegraphics[width=2.0\columnwidth]{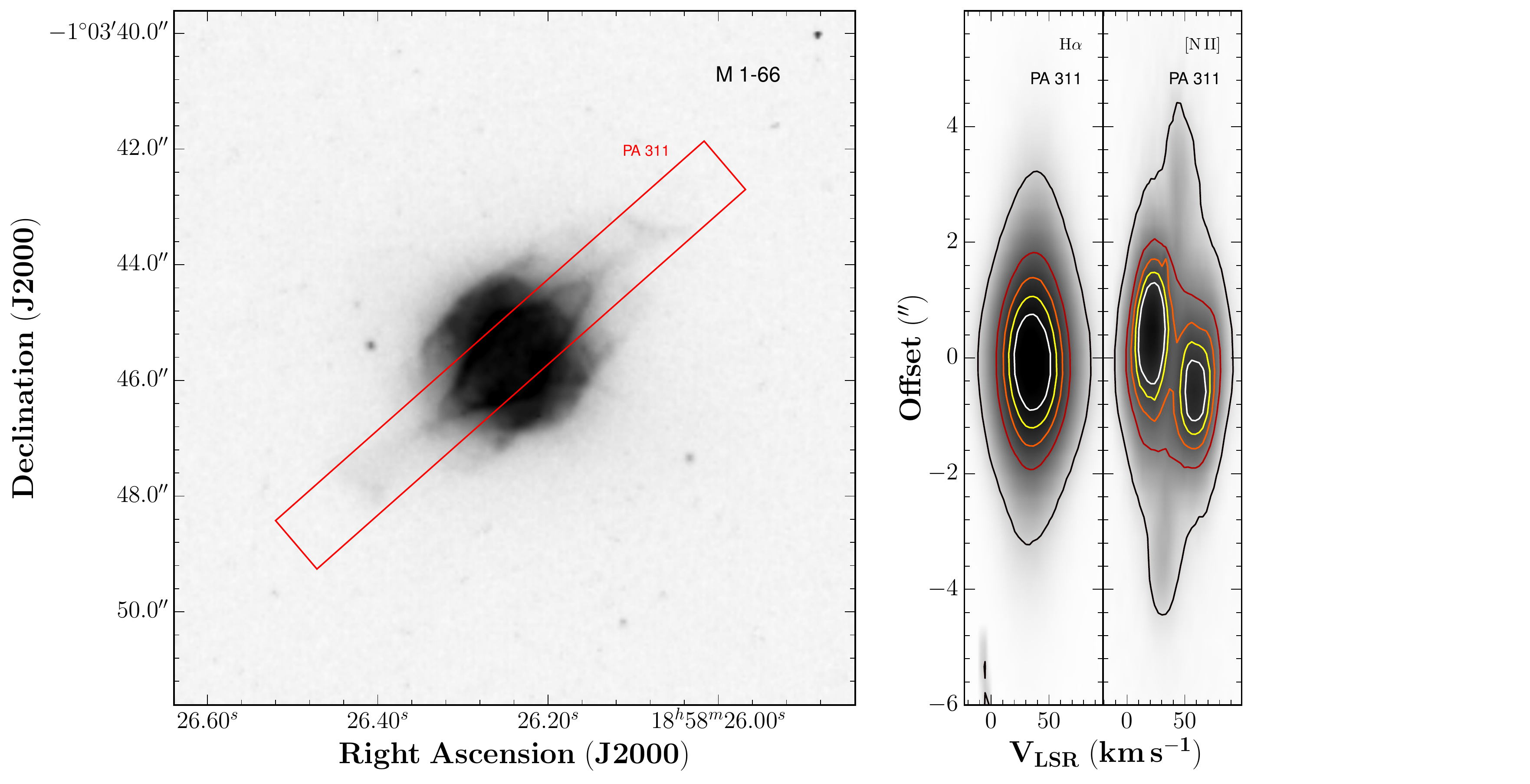}
\caption{
\emph{HST} WFPC2 F656N H$\alpha$ image (left) and H$\alpha$ and [N~{\sc ii}] 
$\lambda$6583 PV maps derived from WHT UES echelle spectra (right) of M\,1-66. 
}
\label{m1-66}
\end{figure*}

M\,1-66 (PN\,G032.7$-$02.0) is a young PN whose rhomboidal inner shell is 
surrounded by an elliptical envelope.  
The first evidence of the presence of a bipolar outflow is found in the 
[N~{\sc ii}] image presented by \citet{MGS96}, later on described as a 
pair of collimated outflow oriented along the major nebular axis at PA = 
131$^{\circ}$ \citep{MG2006}.
These features are more clearly shown in the \emph{HST} image 
presented in the left panel of Figure~\ref{m1-66} \citep{SMV11}.  
A filament departs from the tip of the inner shell and extends up to 
two knots, with a morphology reminiscent of that described for the 
FLIERs of NGC\,7009 \citep{Balick_etal1998}.  
\citet{Socker97} proposed that the progenitor star of M\,1-66 
had interacted with a substellar companion.

We have placed a slit along the pair of collimated outflows 
(PA = 131$^{\circ}$), as shown in the left panel of Figure~\ref{m1-66}.  
The H$\alpha$ and [N~{\sc ii}] $\lambda$6583 PV maps are shown 
in the right panels of this figure. 
A lenticular featureless emission line is observed in the H$\alpha$ PV map, 
but the [N~{\sc ii}] $\lambda$6583 PV map reveals two knots in the equatorial 
regions of the nebula and two faint knots that protrude from the nebula.  
The expansion velocity of the nebular shell is $\simeq$ 20 km~s$^{-1}$.

The [N~{\sc ii}] $\lambda$6583 PV map confirm that M\,1-66 indeed 
harbors a pair of collimated outflows, with very narrow kinematical 
components, although their semi-difference in radial velocity is
small, $v_{\rm r}^{\rm outflow} = \pm$7 km~s$^{-1}$.  
These features have projected distances from the CSPN of 
3.3$\times$10$^{17}$ cm (0.106 pc).

\begin{figure*}
\includegraphics[width=2.0\columnwidth]{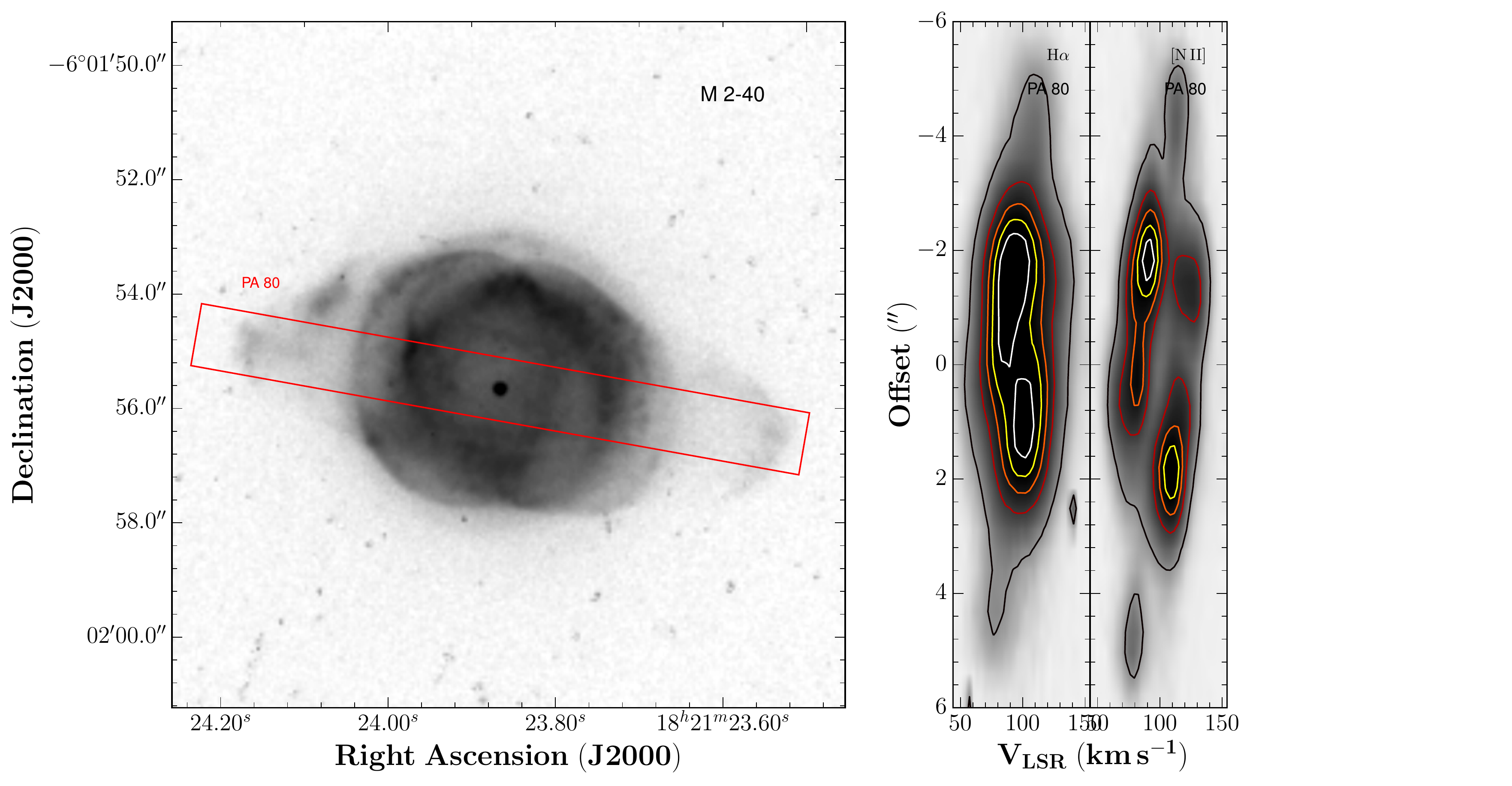}
\caption{
Same as Figure~\ref{m1-66} for M\,2-40.  
}
\label{m2-40}
\end{figure*}

\subsection{M 2-40}

M\,2-40 (PN\,G24.1$+$03.8) is a multiple-shell PN with a detached-halo
interacting with the interstellar medium \citep{GVM98}.
The nebula was classified as elliptical using ground-based images
and the same high-dispersion long-slit data used here \citep{SVM02},
but an \emph{HST} H$\alpha$ image suggests it is bipolar with closed
lobes and major axis along PA = 80$^{\circ}$ \citep{SMV11}.
Two knots protrude from the main nebular shell along the major
axis and produce bow-shock-like features.

The kinematics of  M\,2-40 has been investigated using a slit along 
the major axis at PA = 80$^{\circ}$ (Figure~\ref{m2-40}).  
The H$\alpha$ and [N~{\sc ii}] $\lambda$6583 PV maps reveal a knotty inner
shell with an ellipsoidal rather than bipolar structure.  
Indeed, \citet{GVM98} interpreted this same PV map as evidence
for an ellipsoidal shell.
There is additional faint emission detected at a radial distance of 
$\simeq$10$^{\prime\prime}$ from the CSPN (not shown in the PV maps) 
that corresponds to an arc of enhanced emission resulting from the 
interaction of the halo with the interstellar medium \citep{GVM98}.
The nebula has an expansion velocity $\simeq 17.6$ km~s$^{-1}$,
as derived from the split of the [N~{\sc ii}] emission line at
the location of the CSPN.

\begin{figure*}
\includegraphics[width=2.0\columnwidth]{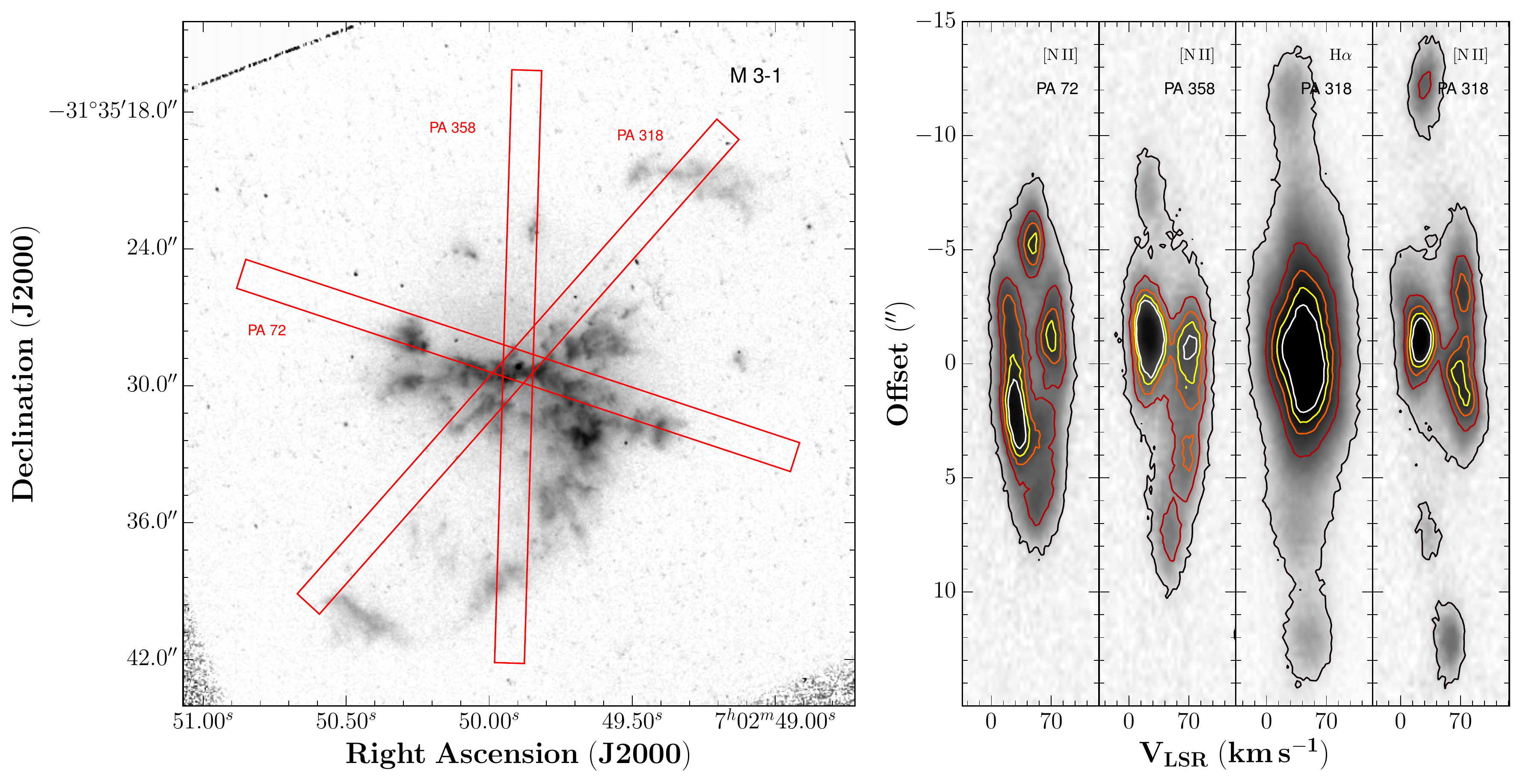}
\caption{
Same as Figure~\ref{hen2-47} for M\,3-1.  
}
\label{m3-1}
\end{figure*}

The pair of knots outside the main nebular shell are detected in these PV maps
and confirmed to be collimated outflows, although their systemic velocity is
not particularly high, $v_{\rm r}^{\rm outflow} = \pm$16.0 km~s$^{-1}$.  
This pair of outflows have projected distances from the CSPN 
$\sim$3.8$\times$10$^{17}$ cm (0.12 pc).

\subsection{M 3-1}

M\,3-1 (PN\,G242.6$-$11.6) is a PN whose morphology has been classified as
point-symmetric \citep{GCM01} based on the H$\alpha$+[N~{\sc ii}] image and 
H$\alpha$+[N~{\sc ii}]/[O~{\sc iii}] ratio map presented by \citet{CMM96}.  
The \emph{HST} [N~{\sc ii}] image (Figure~\ref{m3-1}-left) reveals an 
intricate morphology, with an S-shaped point-symmetric low-ionization 
structure suggestive of a precessing collimated outflow and a filamentary 
disk-like structure at the central nebular regions.  
\citet{BS17} suggested that M\,3-1 \textit{maybe shaped by a triple stellar 
progenitor}.  
It definitely harbors at least a close-binary system with 
a very short period \citep{JBS19}.

The kinematics of M\,3-1 has been investigated using long-slit spectra 
obtained at three slit positions with PA = 72$^{\circ}$ (the disk-like 
structure) and 318$^{\circ}$ and 358$^{\circ}$ (the S-shaped structure).   
The three [N~{\sc ii}] $\lambda$6583 PV maps reveal a knotty 
main body (indicated by the yellow contours in the right panels 
of Fig.~\ref{m3-1}) embedded within a shell of diffuse emission. 
This shell of diffuse emission dominates the H$\alpha$ echellograms, 
as illustrated in the H$\alpha$ PV map along PA = 318$^{\circ}$ 
displayed in the right panel of Figure~\ref{m3-1}.  
The nebula has an expansion velocity $\simeq 24.5$ km~s$^{-1}$.

The S-shaped structure is clearly detected in the PV maps along 
PA = 318$^{\circ}$ and PA = 358$^{\circ}$ as pairs of resolved knots located 
outside the main body, i.e., they are collimated outflows.  
In addition, there is one small knot at PA = 318$^{\circ}$ between the main 
nebular shell and the top knot which correspond to a filament in M\,3-1. 
The systemic velocity of the collimated outflows along
PA = 358$^{\circ}$ is $\pm$14.9 km~s$^{-1}$ and for those
along PA = 318$^{\circ}$ is $\pm$13.2 km~s$^{-1}$.  
The tip of this precessing jet-like structure has projected 
distances from the CSPN of $\sim$8.3$\times$10$^{17}$ cm 
(0.27 pc).

\begin{figure*}
\includegraphics[width=2.0\columnwidth]{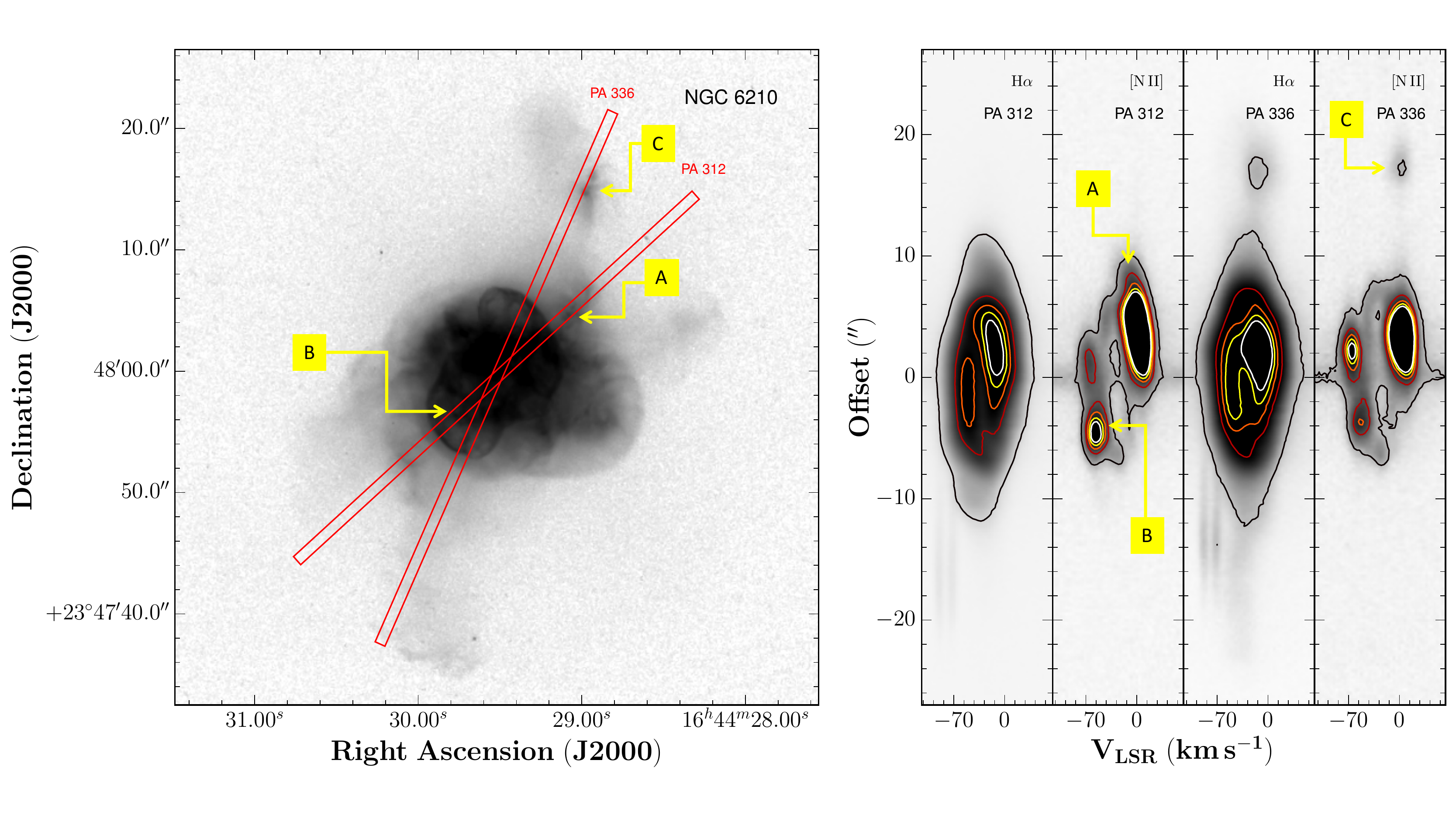}
\caption{
Same as Figure~\ref{hen2-47} for NGC\,6210.  
}
\label{ngc6210}
\end{figure*}

\begin{figure*}
\includegraphics[width=2.0\columnwidth]{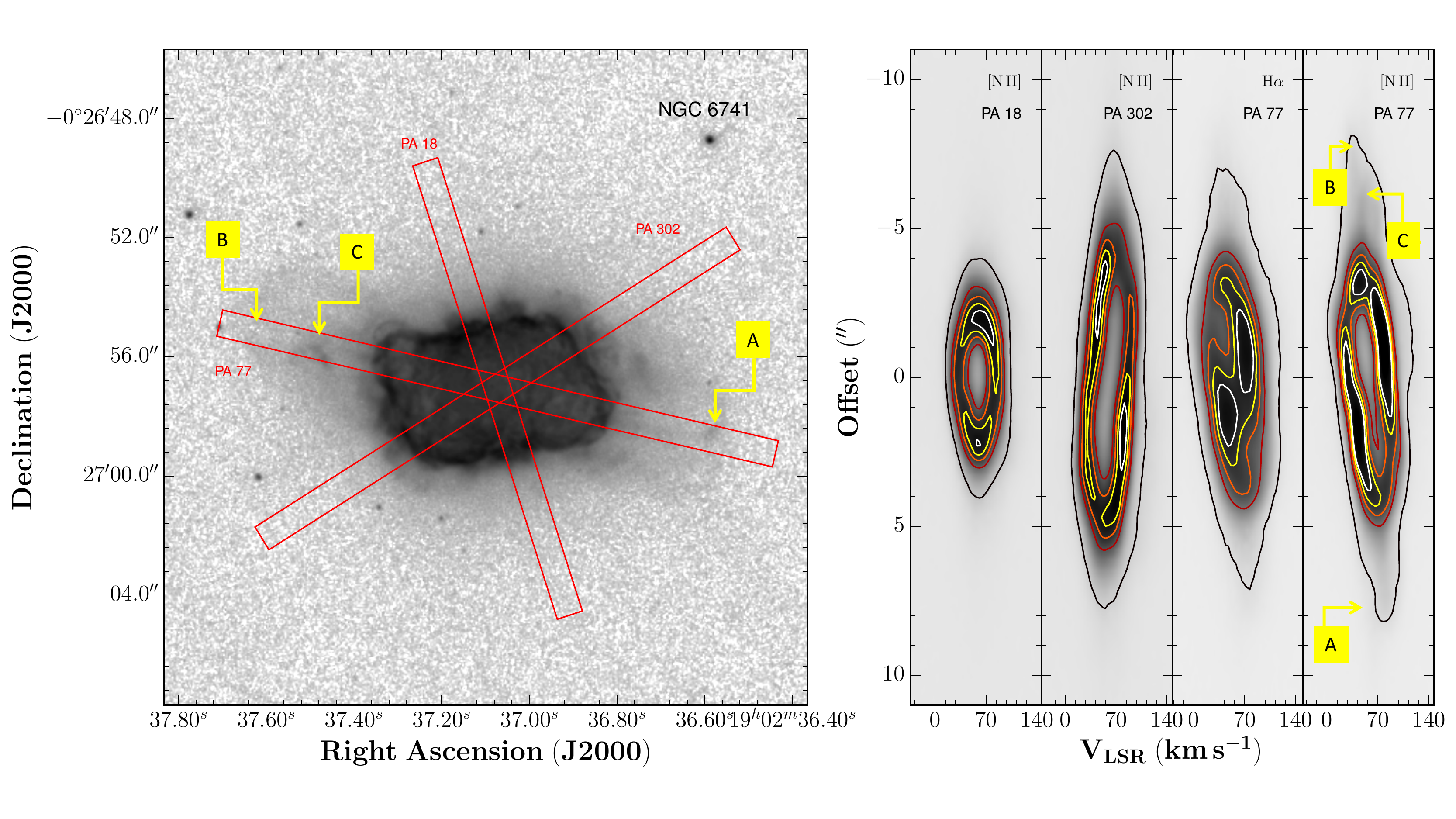}
\caption{
Same as Figure~\ref{hen2-47} for NGC\,6741.  
}
\label{ngc6741}
\end{figure*}

\subsection{NGC 6210}

NGC\,6210 (PN\,G043.1$+$37.7) has a complex shape, described to be amorphous
or irregular \citep{ACB70,Balick87,IPB89} with a general elliptical structure 
\citep{SGCM06,Socker16}.  
This nebula presents a very bright central region with filaments 
and knots \citep{BGF92}, surrounded by a pair of collimated 
outflows with a point-symmetric morphology \citep{PC96,GTM13}.  
These jets are younger that the main shells \citep{GCM01}. 
The nebula is surrounded by a faint halo-like structure \citep{PBR09}.

The kinematics of the complex nebula NGC 6210 was investigated
using two slits.
The first one (PA = 156$^{\circ}$) along a knot located far from the CSPN
labeled as C and also along an extensive outflow with a S-shape fainter
than C.
The second one (PA = 132$^{\circ}$) oriented along two knot-like
features labeled as A and B in Figure~\ref{ngc6210}.
The H$\alpha$ PV maps for both slits show a lenticular shape, while the
[N~{\sc ii}] $\lambda$6583 PV maps show an irregular shape with knots
surrounded by a faint shell.
The expansion velocity of NGC\,6210 is found to be $\simeq 34.2$ km~s$^{-1}$.

The systemic velocity of the collimated outflows along PA = 132$^{\circ}$ is 
$\pm$19.5 km~s$^{-1}$ for outflow A, 
$\pm$30.6 km~s$^{-1}$ for outflow B and, 
for the slit along PA = 156$^{\circ}$, 
$\pm$29.4 km~s$^{-1}$ for outflow C.  
The three outflows labeled as A, B and C have projected 
distances from the CSPN of 
$\sim$2.6$\times$10$^{17}$ cm (0.08 pc), 
$\sim$1.4$\times$10$^{17}$ cm (0.04 pc), and 
$\sim$5.4$\times$10$^{17}$ cm (0.17 pc), respectively.

\subsection{NGC 6741}

NGC\,6741 (PN\,G033.8$-$02.6) has been described as a bright ellipsoidal
nebulae with faint inner structures and filaments based on ground-based
images \citep{Curtis18,ZA86,SCS93b,SBC05}. 
The \emph{HST} [N~{\sc ii}] image of this PN reveals additional faint 
emission outside the main nebular shell and outflow-like features along 
its major axis (Figure~\ref{ngc6741}).

To investigate the kinematics of NGC\,6741, we have placed a slit 
along the outflow-like feature (PA = 77$^{\circ}$) and two along the main 
nebular shell at PA = 18$^{\circ}$ and 122$^{\circ}$. 
All [N~{\sc ii}] $\lambda$6583 PV maps are indicative of a 
shell-like structure (Fig.~\ref{ngc6741}-right), which is 
otherwise not well resolved in the H$\alpha$ PV maps, with
an S-shaped morphology along PA = 77$^{\circ}$.  
In particular, the slit at PA = 18$^{\circ}$ along the minor axis reveals a
toroidal shape, which in conjunction with the PV maps at PA = 18$^{\circ}$
and 122$^{\circ}$ imply that the main nebular shell is tilted along the latter
direction.
Its nebular expansion velocity is $\simeq 23.4$ km~s$^{-1}$.

The PV maps along the major axis reveal the presence of a collimated 
outflow mostly in the [N~{\sc ii}] $\lambda$6583 emission line.  
These are labeled A, B, and C in Figure~\ref{ngc6741}.  
The systemic velocity of features A and B is $\pm$22.7 km~s$^{-1}$,
but only $+$7.3 km~s$^{-1}$ for feature C, suggesting a steep velocity
gradient.  
The pair of outflows A--B have projected distances from the CSPN 
of $\sim$3.9$\times$10$^{17}$ cm (0.13 pc), whereas outflow C is 
found at a projected radial distance of $\sim$3.1$\times$10$^{17}$ 
cm (0.10 pc).

\section{Discussion}

Narrow-band images of a significant fraction of PNe show low-ionization 
morphological features that are suggestive of the presence or action of 
fast collimated outflows.  
Here we have selected a sample of 12 of those PNe and obtained 
kinematical information by means of long-slit high-dispersion 
echelle spectra across morphological features suspected to be 
associated with fast collimated outflows.  
The nature of these features, summarized in the seventh 
column of Table~\ref{tab:result}, can be divided into 
two main groups.

In the first group we include those PNe where indeed the presence of 
collimated outflows is confirmed, namely Hen\,2-429, the outermost 
C-D features of IC\,4776, J\,320, M\,1-66, M\,2-40, M\,3-1, NGC\,6210, 
and NGC\,6741.  
In all those cases, the low-ionization morphological features 
can be associated with narrow kinematical components well 
resolved spatial and kinematically from the main nebular 
shells.  
We note that, for the outermost C-D features of IC\,4776, 
deeper narrow-band images and high-dispersion spectra may 
help eludicate whether these are real collimated outflows 
or a second pair of bipolar lobes.  
Cases of particular interest are those of M\,1-66, M\,2-40, and NGC\,6741, 
whose collimated outflows can be described as jet-like features in the PV 
maps.  
Similarly, the collimated ouflows of Hen\,2-429, M\,3-1, and 
NGC\,6210 are suggestive of precessing jets, whereas those of 
J\,320 can be described as multiple bipolar ejections 
\citep[e.g., Bipolar Rotating Episodic Ejections, BRETs,][]{LVR95}.

In the second group we include those PNe where the presence of fast
collimated outflows cannot be confirmed kinematically, namely
Hen\,2-47, Hen\,2-115, the innermost A-B features of IC\,4776, M\,1-26
and M\,1-37.  
The innermost A-B features of IC\,4776 seem to be located at the tip 
of bipolar lobes, closing a tilted hourglass structure in the PV map. 
The other four PNe share interesting morphological properties, 
with V-shaped lobes in the Starfish Twins Hen\,2-47 and M\,1-37, 
and in Hen\,2-115 and M\,1-26.  
The PV maps along these features imply kinematics differing from those 
of their main nebular shells, particularly for Hen\,2-47 and M\,1-26, 
with notably higher expansion velocities.  
These perturbations in the kinematics, together with the peculiar morphology 
of these features, may result from the action of stellar outflows arising at 
the end of the AGB phase \citep{ST98} that we cannot detect either because 
they are too tenuous or because they are not ionized, as the molecular 
outflows reported in NGC\,7027 \citep{Cox_etal2002,Huang_etal2010}.  
Apparently, these outflows have not been able to pierce through the
nebular envelope, which has retained the linear momentum from the
outflow and therefore experienced a notable acceleration.  
This points out to ``light'' outflows, defined to have densities 
much lower than that of the slow AGB wind \citep{AS2008}.

The velocity signature of these V-shaped morphological features can be 
investigated using their apparent opening angle to estimate the Mach 
number of the ejections producing these lobes.  
This is referred as the Mach angle $\alpha$, and can be related to the Mach
number $M$ as
\begin{equation}
\sin \alpha = \frac{1}{M},
\end{equation}
\noindent 
where $M$ is defined as the ratio $v/c_\mathrm{s}$ between the velocity
of the Mach front $v$ and the isothermal sound speed of ionized gas
$c_\mathrm{s}\approx10$~km~s$^{-1}$. 
The Mach angles derived for Hen\,2-47, Hen\,2-115 and M\,1-37 are 
27$^{\circ}$, 23.5$^{\circ}$, and 25$^{\circ}$, which correspond to 
very similar Mach numbers of 2.20, 2.50, and 2.36, respectively.  
As for M\,1-26, no Mach number has been derived because 
its lobes do not have well defined V-shape.  
For inclination angles with the plane of the sky smaller than 45$^\circ$, 
these expansion velocities can be added quadratically with the radial 
velocities listed in Table~\ref{tab:result} to derive the space velocity 
of the ejections that produced these V-shaped lobes with an uncertainty 
caused by the unknown inclination angle smaller than 10\%.  
These are 35~km~s$^{-1}$, 25~km~s$^{-1}$, and 24~km~s$^{-1}$ 
for Hen\,2-47, Hen\,2-115 and M\,1-37, respectively. 
Therefore, the V-shaped lobes of these sources expand at velocities 
$\sim$2--3 times that of their (slow) main nebular shells (see 
Table~\ref{tab:result}).

It is interesting to note the similar morphology of the V-shaped
structure along the major axis and direction of the collimated outflow
of Hen\,2-429 and particularly M\,1-66.  
This may indicate that the collimated outflows in these two PNe had 
sufficient momentum to break through the nebular envelope from the 
AGB. 
In this sense, the collimated outflows in NGC\,6543 and NGC\,7009, 
arising from V-shaped features at the tip of the major axis of their 
inner shells, can be interpreted as much more evolved versions of 
these interactions.  
It can thus be envisaged an evolutionary sequence where 
collimated outflows start piercing through the nebular envelope 
like in Hen\,2-47, Hen\,2-115, M\,1-26, and M\,1-37, then break 
outside this envelope like in Hen\,2-429 and M\,1-66, and finally 
emerge like in NGC\,6543 and NGC\,7009.  
The kinematical ages of PNe with and without collimated outflows have 
similar ranges, and thus they do not support this interpretation,
although uncertainties in their distances and assessment of their true 
ages through the kinematic age may hinder such trend \citep{SJS2005}.  
There is additional evidence that the four PNe without clear 
collimated outflows belong to a different group from the other 
PNe in our sample.  
All of them can be classified as very low excitation PNe 
([O~{\sc iii}]/H$\alpha \lesssim$1) with low effective 
temperature CSPNe \citep{ST98,SMV11}, which can be 
interpreted as a sign of either an early evolutionary phase 
or slow post-AGB evolution as expected for smaller mass 
progenitors \citep{MM2016}.
It is interesting to note that these sources also exhibit the smallest
nebular expansion velocities among our sample of PNe (Tab.~\ref{tab:result}),
which would also place them at an early evolutionary phase 
\citep{RLP_etal2008,RLG_etal2010} according to theoretical models 
of nebular evolution \citep{JSS2013}, but that can also be 
explained if their progenitors had low masses \citep{PLR2016}.

\section{Summary}

We present a spatio-kinematical investigation of a sample of twelve PNe 
suspected to have collimated outflows using archival narrow-band images 
and high-dispersion long-slit echelle spectra.  
Collimated outflows, which can be described as distinct narrow kinematical 
components with notable velocity variations with respect to the main nebular 
shells associated with compact knots and linear or precessing jet-like 
features, are confirmed in seven sources, with one additional possible 
detection in IC\,4776.  
In all other cases, the nature of the possible collimated outflows 
is not confirmed or even rejected, as the case of the innermost 
A-B components of IC\,4776, which can be associated with a pair of 
bipolar lobes.  
The morphology and kinematics of the four sources where collimated outflows 
are not confirmed, namely the Starfish Twins Hen\,2-47 and M\,1-37, and 
Hen\,2-115 and M\,1-26, are indicative of the action of collimated outflows 
which are not detected in optical images and spectra. 
These PNe appear to be at an early evolutionary phase or 
descend from lower mass progenitors than the PNe in our 
sample with unambiguous collimated outflows.

\section*{Acknowledgments}

JSR-G and MAG acknowledge support of the grants AYA\,2014-57280-P and 
PGC2018-102184-B-I00, co-funded with FEDER funds.
SDP acknowledges financial support from the Spanish Ministerio de
Econom\'\i a y Competitividad under grants AYA2013\,47742-C4-1-P
and AYA2016\,79724-C4-4-P, from Junta de Andaluc\'\i a Excellence
Project PEX2011\,FQM-7058.
JAT and MAG are also funded by UNAM DGAPA PAPIIT project IA100318.  
LFM acknowledges partial support by grant AYA2017-84390-C2-R, co-funded 
with FEDER funds. 
JSR-G, MAG, LFM, and SDP acknowledge financial support from the State Agency
for Research of the Spanish MCIU through the "Center of Excellence Severo
Ochoa" award for the Instituto de Astrof\'\i sica de Andaluc\'\i a
(SEV-2017-0709).  
We thank Dr.\ Laurence Sabin for a thorough reading of 
the manuscript and her valuable comments that helped 
us to improve it.

Based on observations made with the NASA/ESA
\emph{Hubble Space Telescope}, and obtained from the \emph{Hubble}
Legacy Archive, which is a collaboration between the
Space Telescope Science Institute (STScI/NASA), the
Space Telescope European Coordinating Facility (STECF/
ESA) and the Canadian Astronomy Data Centre
(CADC/NRC/CSA).
This research has made also use of the NASA's Astrophysics Data System.

This research made use of Python 
({\tt \href{http://www.python.org}{http://www.python.org}}) 
and IPython \citep{PER-GRA:2007}, APLpy \citep{2012ascl.soft08017R}, 
Numpy \citep{2011arXiv1102.1523V}, Pandas \citep{mckinneyprocscipy2010}, 
and Matplotlib \citep{Hunter:2007}, a suite of open-source Python modules 
that provides a framework for creating scientific plots. 
This research made use of Astropy, a community-developed core Python 
package for Astronomy \citep{2013A&A...558A..33A}. 
The Astropy web site is 
{\tt \href{http://www.astropy.org/}{http://www.astropy.org}}.

\bibliographystyle{mn2e}

\bsp


\end{document}